\documentclass[journal]{IEEEtran}
\usepackage{graphicx}
\usepackage{amsmath}
\usepackage{enumerate}
\usepackage{multirow}
\usepackage{array}
\usepackage{float}
\usepackage{subfigure}
\usepackage{epstopdf}
\usepackage{balance}
\usepackage[noadjust]{cite}
\usepackage{array,booktabs}

\newcolumntype{C}{>{$\displaystyle}c<{$}}
\ifCLASSINFOpdf
 \else

\fi

\begin{document}
\title{A Survey on Human-centric Communications in Non-cooperative Wireless Relay Networks}

\author{Behrouz Jedari, Feng Xia, \IEEEmembership{Senior Member, IEEE}, and Zhaolong Ning% <-this % stops a space
\thanks{The authors are with the Key Laboratory for Ubiquitous Network and Service Software of Liaoning Province, School of Software, Dalian University of Technology, Dalian 116620, China.}% <-this % stops a space
\thanks{Corresponding author: Feng Xia; E-mail: f.xia@ieee.org}% <-this % stops a space
\thanks{Manuscript received XX XX, 2017; revised XX XX, 20XX.}}

\markboth{IEEE Communications Surveys \& Tutorials, Vol. XX, No. XX, 2018}%
{Shell \MakeLowercase{\textit{et al.}}: Bare Demo of IEEEtran.cls for Journals}

\maketitle

\begin{abstract}
The performance of data delivery in wireless relay networks (WRNs), such as delay-tolerant networks and device-to-device communications heavily relies on the cooperation of mobile nodes (\textit{i.e.}, users and their carried devices). However, selfish nodes may refuse to relay data to others or share their resources with them due to various reasons, such as resource limitations or social preferences. Meanwhile, misbehaving nodes can launch different types of internal attacks (\textit{e.g.}, blackhole and trust-related attacks) to disrupt the normal operation of the network. Numerous mechanisms have been recently proposed to establish secure and efficient communications in WRNs in the presence of selfish and malicious nodes (referred as non-cooperative WRNs). In this paper, we present an in-depth survey on human-centric communication challenges and solutions in the non-cooperative WRNs that focuses on: (1) an overview of the non-cooperative WRNs and introduction to various types of node selfish and malicious behaviors, (2) the impact analysis of node selfish and malicious behaviors on the performance of data forwarding and distribution, (3) selfish and malicious node detection and defense systems, and (4) incentive mechanisms. Finally, we discuss several open problems and future research challenges.
\end{abstract}


\IEEEpeerreviewmaketitle

\section{Introduction}

\noindent Today, individuals primarily use their handheld devices, such as smartphones and tablets for daily business communications and entertainment (\textit{e.g.}, mobile advertising, file sharing, and gaming), which leads to exploding traffic over mobile networks. The global cellular traffic reached 7.2 exabytes per month at the end of 2016, and it is expected to grow to 49 exabytes per month by 2021 \cite{CISCOReport2016}. Thus, it has become a great challenge for the Internet providers and mobile network operators to serve the booming traffic demand of cellular networks. Meanwhile, mobile users in emergency scenarios may not have access to the Internet due to some reasons, such as limited coverage of cellular networks (\textit{e.g.}, 3G or LTE). To overcome these problems, wireless relay networks (WRNs) have emerged as a promising communication paradigm in which the architecture of delay-tolerant networks (DTNs) \cite{DTNFall-142}\cite{DTNBOokVasilakos} is incorporated to establish device-to-device (D2D) communications \cite{D2DSurveyAsadi} between mobile nodes (\textit{i.e.}, users and their devices). In WRNs, nodes in proximity can opportunistically communicate and share their resources with each other using short-range and high-speed wireless interfaces, such as Wi-Fi and LTE Direct, which can significantly reduce the traffic of the cellular network. For instance, mobile social networks (MSNs) \cite{MSNSury-124} have emerged as a novel networking paradigm in WRNs wherein the nodes' social relationships and contextual information are leveraged to enhance their communications and improve the resulting network performance.
WRNs have many applications in different areas, such as mobile data offloading \cite{OppDataOffloading}, proximity services \cite{D2DProximityServices}, public safety communications \cite{D2DPublicSafety}, and vehicular networks \cite{VehicularDTNs}\cite{VehicularNet}.

The primary goal of data forwarding and sharing protocols in WRNs is to exploit the nodes' contact, context, and social information to improve the data delivery performance in terms of different metrics (\textit{e.g.}, delivery ratio, delay, overhead, and energy consumption). The majority of existing protocols assume that mobile nodes willingly participate in data delivery, share their resources with each other, and follow the rules of underlying networking protocols. Nevertheless, rational nodes in real-world scenarios have strategic interactions and may exhibit selfish behaviors due to various reasons (such as resource limitations, the lack of interest in data, or social preferences). For example, in case a node has limited battery resources or the cost of the network bandwidth delivered by mobile network operators is high, it would not be willingly to relay data for others until appropriate incentives are provided. Meanwhile, malicious nodes may attack the network in different ways to disturb the normal operation of the data transmission process. An adversary, for example, may drop received messages
%\footnote{We note that a message in the context of this paper denotes a data packet, while a message in different research areas may have different meanings.}
but produce forged routing metrics or false information with the aim of either attracting more messages or decreasing its detection probability. This issue becomes more challenging when some colluding attackers boost their metrics to deceive the attack detection systems. However, dealing with the non-cooperative behaviors of mobile nodes in WRNs is very challenging because of the distributed network model and intermittent node access to centralized authorities.

Recently, extensive analytical and simulation-based experiments have been conducted to study the effects of mobile nodes' selfish and malicious behaviors on the performance of data forwarding and dissemination in DTNs and D2D communications underlying cellular network. Besides, several distributed algorithms have been proposed to detect the nodes' selfish and malicious behaviors and protect the network against malicious attacks. Furthermore, a large number of incentive mechanisms, such as reputation and rewarding approaches have been developed in both DTNs and D2D communications to either exclude selfish nodes from the data delivery process or stimulate them to participate in data relaying.

\subsection{Prior Related Surveys}
\noindent In the past few years, some survey articles have been presented in the context of WRNs. The majority of existing studies address the design requirements, platforms, and applications of DTNs and MSNs \cite{MSNSury-124,MSNSurvey-126,MSNSur-127,MSNSur-128,MSNPWang2014}. For instance, Kayastha \textit{et al.} \cite{MSNSurvey-126} categorize MSNs into two types: infrastructure-based and infrastructure-less (or opportunistic) and discuss their architectures and characteristics. A number of studies review data routing and dissemination protocols in DTNs and MSNs \cite{MSNSur-129,OppRouSur-130,OppRou-131,Spyropoulos2010,OppRoutingBook1} and categorize them into different classes according to various factors (\textit{e.g.}, contact, context, and social features). Youssef \textit{et al.} \cite{RoutingMetrics} explore different routing metrics, and Abdelkader \textit{et al.} \cite{DTNPerf-132} evaluate the performance metrics of some well-known opportunistic routing protocols. Zhu \textit{et al.}\cite{PosNeg-133} study the positive (\textit{e.g.}, social similarity and centrality) and negative (\textit{e.g.}, user selfishness) aspects of data delivery algorithms in MSNs. The authors in \cite{MobImpRoutSur-134}\cite{PIROZMAND201445} study the impact of human mobility on the performance of opportunistic routing protocols. The authors in \cite{UserBehaviorOSNs2013,HumanIntTemporal2015,LI2015383} respectively explore human behavior in social, temporal, and microblog networks. Silva \textit{et al.} \cite{CopStrSur-135} study different cooperative strategies and their applications in challenged networks. The authors in \cite{MobOffliadingSur}\cite{UserProvided}\cite{Contract3} study design challenges of incentive strategies and their trade-offs for data forwarding in wireless networks. Furthermore, Ahmed \textit{et al.} \cite{Event-basedMSN} study the services, technologies, and applications of event-based MSNs.

Recently, some articles study recent advances in D2D communications. Asadi \textit{et al.} \cite{D2DSurveyAsadi} classify D2D communications into in-band and out-band, \textit{i.e.}, communication on the cellular and unlicensed spectrum, respectively, where the main difference is the interference caused by D2D nodes. Wang \textit{et al.} \cite{D2DMSN2017} investigate the key components and architecture of D2D-based proximity services in MSNs and highlight their challenges and existing solutions. Zhao and Song \cite{DTN-D2DSurvey} provide an overview of social-aware data dissemination approaches in MSNs and D2D communications with respect to game theory, matching theory, and optimization techniques. Gandotra \textit{et al.} \cite{GANDOTRA20179} study the implementation challenges of D2D communications from several aspects, such as resource allocation and interference management. In addition, Ahmed \textit{et al.} \cite{SocialD2DResAlloc2} study resource allocation approaches in social-aware D2D communications with respect to their channel information, communication type, and networking technologies.

%Li and Guo \cite{IncentiveD2D2015} study incentive mechanisms in D2D communications.

A couple of survey articles have explored security aspects of human-centric communications in WRNs. Najaflou \textit{et al.} \cite{SafetyMSN-136} study safety challenges in MSNs in three main groups: security, trust, and privacy. Liang \textit{et al.} \cite{SeuPriSur-137} provide a brief overview of MSN applications with respect to security and privacy and highlight some future research challenges about secure routing and denial-of-service attacks in MSNs. Furthermore, Zhang \textit{et al.} \cite{SybilSur-138} study various types of Sybil attacks and their defense mechanisms in a broad context of wireless networks. Haus \textit{et al.} \cite{D2DSecuPriSurvey} present a survey on privacy and security in D2D communications. Despite the fact that the existing studies have outlined different aspects of WRNs, there is no prior in-depth survey of communication challenges and solutions in non-cooperative WRNs.

\subsection{Contributions of this Survey}

\noindent To the best of our knowledge, this paper is the first survey that provides a comprehensive review of existing work on human-centric communications in non-cooperative WRNs. Our major contributions can be summarized as follows:

\begin{itemize}
\item We present an overview of non-cooperative WRNs and introduce mobile nodes' different selfish behaviors and attack models.
\item We survey recent studies that explore the impact of nodes' selfish and malicious behavior on the performance of data forwarding and distribution protocols in WRNs and explore their detection and defense mechanisms.
\item We study numerous incentive mechanisms in WRNs and discuss their important characteristics.
\item We discuss several open issues and highlight future research directions regarding data forwarding and distribution in non-cooperative WRNs.
\end{itemize}

\subsection{Methodology}
The main goal of this survey is to provide a structured and comprehensive overview of human-associated communications in non-cooperative WRNs. In particular, we explore data delivery in proximity-based networks under the circumstances that some mobile nodes exhibit selfish and malicious behavior to either maximize their utility or disrupt the data delivery process. In Section II, we present an introduction to non-cooperative WRNs with the aim of motivating the emergence of protocols and mechanisms to deal with non-cooperative behaviors in WRNs. Furthermore, we outline different forms of nodes' selfish and malicious behavior in data forwarding. Next, we study data delivery challenges and solutions in non-cooperative WRNs from three perspectives. In Section III, we study proposals that analyze the impact of nodes' selfish behaviors on the performance of data delivery protocols. In particular, we categorize existing methods into simulation-based, theoretical, and hybrid methods and highlight their principal solutions, specialties, and limitations in Table \ref{tableImpact}.

\begin{figure*}[!t]
\centering
\includegraphics[width=5.5 in]{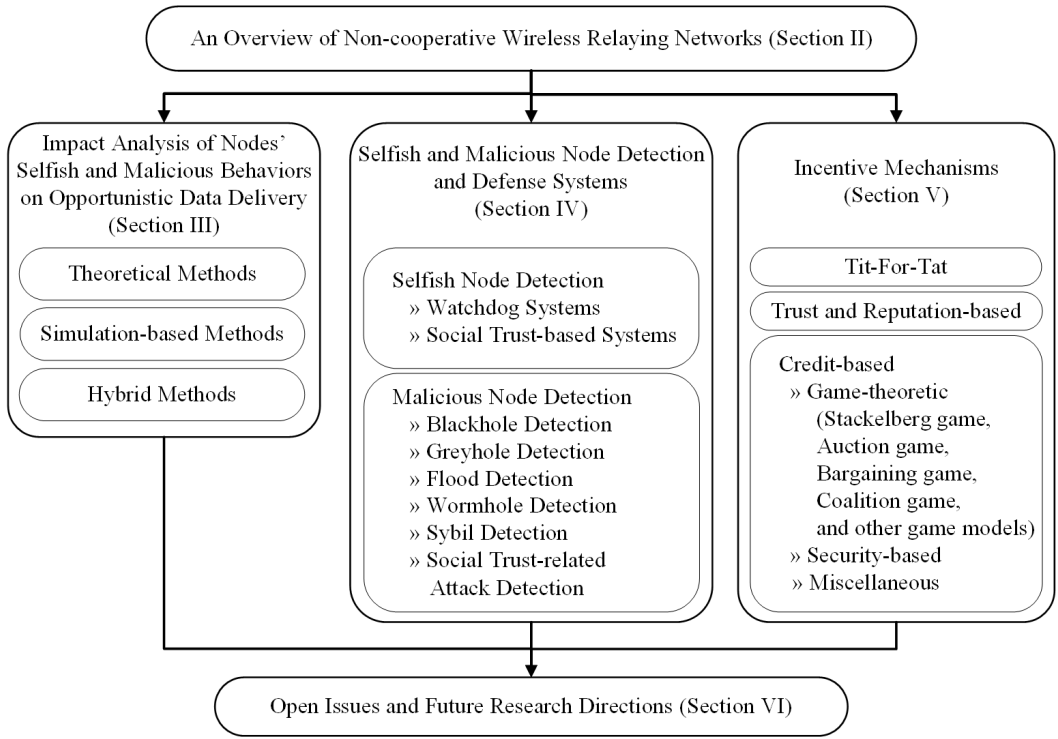}
\caption{The organization of the remaining parts of the paper.}
\label{Outline}
\end{figure*}

In Section IV, we study selfish and malicious node detection and isolation algorithms in WRNs where they aim to secure data delivery protocols against nodes' non-cooperative behaviors. We categorize the selfish node detection and isolation methods into two classes (watchdog systems and social trust-based systems) and highlight their major contributions, properties, and limitations in Table \ref{tableWatchdog}. Next, we explore proposals that aim to detect different types of node attacks in WRNs and outline their principal contributions, major properties, and shortcomings in Table \ref{tableAttackDetect}. In Section V, we study incentive mechanisms that aim to promote the cooperation of nodes in data relaying where we categorize existing methods into three main classes: Tit-For-Tat, reputation-based, and credit-based mechanisms. First, we introduce prominent Tit-For-Tat and credit-based schemes and outline their major characteristics in Table \ref{TFTReputationIncentives}. Next, we classify credit-based mechanisms into three classes (game-theoretic, security-based, and miscellaneous) and highlight their principal solutions, incentive objects, and limitations in Tables \ref{SecurityMiscellaneouscredits} and \ref{IncentiveGame}. We believe that this paper can educate the research community and networking protocol designers how to effectively deal with non-cooperative behaviors of mobile carriers in next-generation wireless networks.

The rest of the paper is organized as follows (Fig. \ref{Outline}). Section II provides an overview of non-cooperative WRNs and introduces various selfishness and attack models. Section \ref{Impact} introduces different approaches that study the impacts of node selfish and malicious behavior on the performance of data forwarding and sharing protocols in WRNs. Section \ref{Detection} discusses the selfish node detection techniques and attack defense systems. Section \ref{Incentive} studies representative incentive mechanisms that aim to either promote the cooperation of selfish nodes or exclude them from the data delivery process. Section \ref{FutureWork} provides several open problems and future research directions, and Section \ref{Conclusion} draws the conclusion.

\section{An Overview of Human-centric Communications in Non-cooperative WRNs}
\label{Overview}

\begin{figure*}[!t]
\centering
\includegraphics[width=6.1 in]{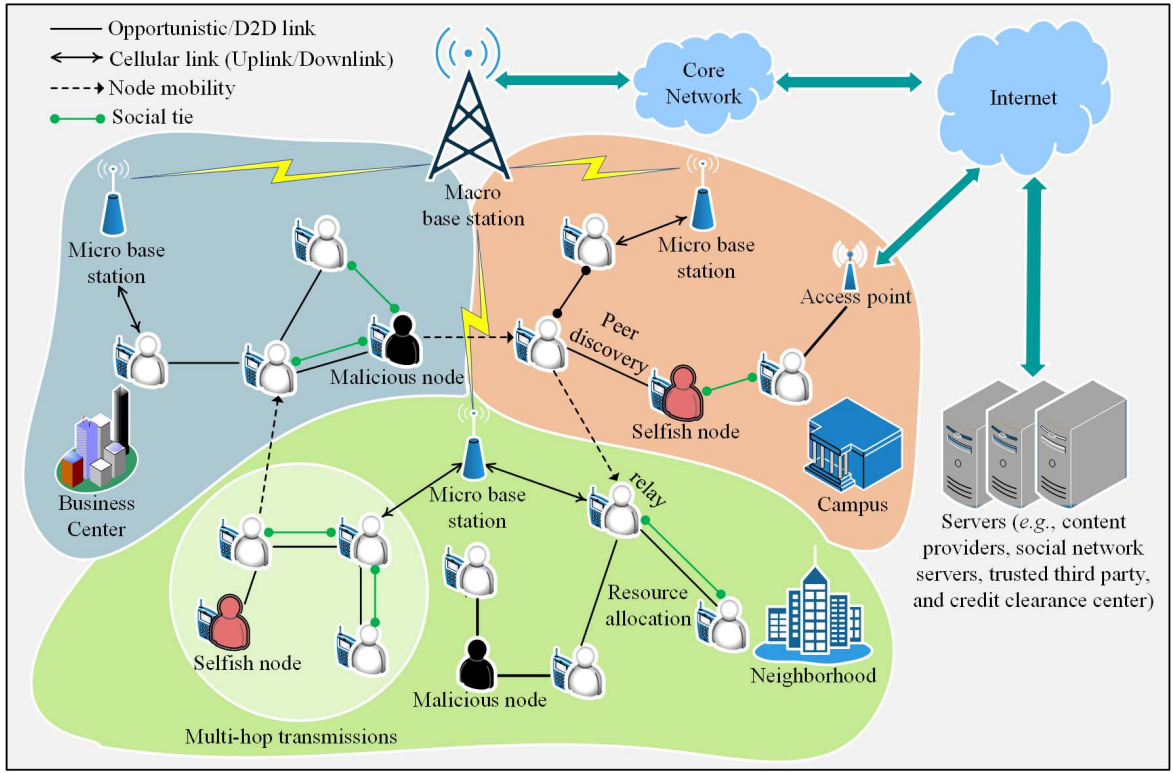}
\caption{An overview of non-cooperative wireless relay networks.}
\label{OverviewFig}
\end{figure*}

\noindent Fig. \ref{OverviewFig} illustrates an overview of human-centric communications in non-cooperative WRNs. As shown in the figure, mobile nodes (or user equipments) in proximity can establish peer-to-peer communications to exchange data with each other using short-range and high-speed wireless transmission technologies (such as Bluetooth, ZigBee, WiFi-Direct or LTE Direct) \cite{ShortRangeInter}. The communication between the nodes can be in standalone D2D mode (or ad hoc mode) autonomously or via network-assisted D2D communications with the control of base stations (BSs) or core network. Meanwhile, the nodes may sporadically have access to the Internet and service providers (such as a trusted third party or credit clearance center) via BSs or Wi-Fi hotspots. In this setting, the nodes' social ties and relationships captured from their online social network profiles or the nature of their mobility (\textit{e.g.}, contact patterns or geographic information) can be leveraged to enhance their communications and capacity of the network.

Cooperative communications can improve the performance of data delivery in WRNs and offload the traffic of the cellular network. For example, mobile devices in D2D communications can cache popular content received from the cellular network and share them with interested neighbor requestors, which can improve the data delivery performance, increase the network capacity, and offload the traffic of BSs. However, some nodes might exhibit selfish behavior and refuse to relay messages received from all or some other nodes or share their resource with them because of different reasons, such as limited resources (\textit{i.e.}, buffer, bandwidth, and energy resources) or monetary cost. In addition, malicious nodes can launch different forms of attacks, such as manipulating and diffusing wrong information to deceive the nodes and disturb their normal communications. Thus, we can classify mobile nodes into three types: cooperative, selfish, and malicious nodes. In general, the cooperative nodes follow the rules of the underlying networking protocols, whereas the selfish nodes consume the network resources but refuse to provide services for all or some other nodes with the aim of maximizing their own benefits. Besides, the malicious nodes attack the network in different ways to disrupt the network normal functionalities. In the rest of this section, we explain social-awareness communications in WRNs and compare the characteristics of DTNs and D2D communications. Next, we discuss different selfish behaviors and malicious attacks that can be launched by non-cooperative nodes in WRNs.

\subsection{Social-aware Communications in WRNs}

Mobile nodes' implicit (\textit{e.g.}, mobility information) and explicit (\textit{e.g.}, online social network information) social characteristics and relationships can accurately mirror their interactions and relationships in the real life. Hence, socially-aware wireless networking has emerged as a promising solution to optimize various aspects of human-centric communications in WRNs \cite{SANParadigm-125,SocialReplic-Adhoc,BEEINFO,Geo-Soc,Soc-Oriented}. In particular, nodes' different social characteristics, such as social ties, community, and centrality are primarily exploited to enhance different key technological problems in cooperative communications underlying cellular network \cite{D2DCooperation2017}\cite{YongLiSocialD2D2014}. For instance, nodes' social network and mobility information can be leveraged to select appropriate relay nodes in D2D communications with the aim of improving the data delivery success ratio while minimizing the communication overhead (\textit{e.g.}, see \cite{Social-awareD2D} and \cite{FusingSocial-115}). In addition, nodes' social characteristics are exploited to address peer discovery \cite{SocialD2DPeerDiscov} and resource allocation \cite{SocialD2DResAlloc1} in D2D communications.

In contrast to cooperative networks, the social information of mobile nodes can also be exploited to achieve secure communications in non-cooperative WRNs. For instance, social-based trust or reciprocity relationships between interacting parties can streamline data delivery performance and protect their communication against malicious attacks (\textit{e.g.}, see\cite{MILITANO2016141}\cite{OMETOV2016327}). In addition, the social features and behaviors of nodes can help detect their possible selfish and malicious actions \cite{D2DSelfMaliDetect}. Furthermore, utilizing nodes' social features can help model their interactions and incentive mechanisms realistically with respect to their similar and conflicting interests \cite{SocialRate2017}. In this paper, we particularly study proposals that leverage nodes' social attributes and relationships across different aspects of their communications in non-cooperative WRNs.

\subsection{DTNs and D2D communications}
Although mobile carriers in both DTNs and D2D communications can establish opportunistic contacts to exchange their messages, there are some distinct differences in the form of their communications. Typically, there is no permanent cellular infrastructure in DTNs and the research question is how to efficiently deliver a message from a source node to its destination node by choosing appropriate relay nodes. In contrast, the main goal in D2D communications is to efficiently offload the traffic distributed by a cellular network through D2D devices to interested nodes, which is applicable in new business models and scenarios (\textit{e.g.}, pervasive social networks and location-based services). In other words, there is no strict publish/subscribe model in DTNs in comparison with the data offloading mechanisms assumed in D2D communications. In addition, DTNs primarily employ multi-hop relaying to deliver messages to destination nodes, whereas D2D communications apply single-hop or multi-hop cluster-based transmissions. Furthermore, mobile nodes in DTNs communicate with each other on an unlicensed spectrum, which is performed by the devices autonomously. In contrast, D2D devices can use both licensed and unlicensed spectrum under the controlled of the BS or within the cooperation between the BS and encountered nodes, which can cause D2D-to-cellular and cellular-to-D2D interference \cite{D2DSecSurvey}. Hence, resource allocation, peer discovery, mode selection, and power management are major challenges in D2D communications \cite{CoopD2D2015}.

\subsection{Node Selfishness Models}

\noindent Although cooperation among mobile nodes in proximity can improve the data delivery performance in WRNs, some nodes may exhibit selfish behavior and do not share their resources with other nodes altruistically with the aim of maximizing their preferred utility. The selfish behavior of mobile carriers could have different reasons, such as resource constraints, the lack of interests in messages, privacy concerns, or social preferences. For example, in case a mobile node has limited battery resources or the cost of network bandwidth delivered by mobile network operators is high, it may not be willing to consume its resources and relay data for all or some other nodes until appropriate incentives are provided.

Different forms of node selfishness models have been considered in the literature. The authors in \cite{Selfishalturisim-14} propose different altruism distributions, such as uniform, degree-biased, and community-biased to realize human selfish behaviors in WRNs. Some studies identify a \textit{probabilistic selfish behavior} in which a selfish node may not participate in relaying a message according to a probabilistic function. A number of studies (\textit{e.g.}, \cite{EffNonCoo-15}) define \textit{non-forwarding} and \textit{partially-forwarding} selfish actions where a selfish node does not relay messages to other nodes or only delivers the relaying messages to their destination nodes. Panagakis \textit{et al.} \cite{Effect-12} introduce \textit{non-copying} (or dropping) and \textit{non-forwarding} selfish behavior. In addition, the authors in \cite{VirtueSelfishness-16} define two types of selfish nodes: strict and mild. A strict selfish node turns off its radio interface after receiving its requested data items, whereas a mild selfish node cooperates with others for a limited time even after receiving its requested data. Besides, the authors in \cite{PerfModEpid-23} introduce egotistic nodes, which change the range of their communication signals in different situations.

Despite various selfishness models and actions mentioned above, a vast number of existing studies in non-cooperative WRNs have explored the role of nodes' social relationships and preferences in their selfish behavior. Following the homophily phenomenon in sociology, it is revealed that mobile nodes usually provide better services for those with whom they have strong social relationships or similarities. For example, nodes with similar interests and backgrounds tend to cooperate with each other in data delivery, even if they have not had direct contact with each other previously \cite{SSAR-120}\cite{SimFavor-119}. Thus, two types of selfish nodes can be defined as follows:

\begin{itemize}
\item \textbf{Individually Selfish (IS) nodes:} IS nodes have socially-oblivious selfish behavior and exhibit a uniform selfish behavior to other nodes without considering the utility of the nodes with whom they have social relationships or common interests. For example, an IS node does not consider the benefits of its friends in data sharing and provides better services for nodes with early access times.

\item \textbf{Socially Selfish (SS) nodes:} SS nodes alleviate the degree of their selfishness degree based on their social relationships or similarities to provide better services to their friends or nodes with whom they have strong social ties. In contrast, they are unwilling to provide forwarding services for strangers or nodes with different social objectives or preferences with the aim of saving their buffer and energy resources. For example, SS nodes in community-based DTN or D2D data offloading scenarios are willing to cache and deliver messages to nodes in the same community but refuse to relay the messages to nodes in other communities.
\end{itemize}

\subsection{Social Trust}
Social trust is a powerful descriptor of friendship, honesty, security, and integrity that can secure interactions between mobile nodes in wireless networks. In particular, due to the lack of a permanent central authority in WRNs, establishing social trust relations between nodes (by leveraging their online social network information, direct, and indirect interactions) can promote trustworthy cooperation among them and protect them against threats and attacks \cite{TrustSurvey}. For example, social trust can improve the performance of D2D communications by asking the most trustworthy nodes in proximity (\textit{e.g.}, family members, friends, or colleagues) to relay messages \cite{TrustedSocD2DOmetov}. In contrast, the lack of trust can make the nodes reluctant to cooperate with each other due to different reasons ranging from privacy concerns (\textit{e.g.}, not trusting to interact with strangers) to resource constraints (\textit{e.g.}, energy and buffer limitations). However, malicious nodes can attack the trust system, for example, by exaggerating the reputation of other malicious nodes or submitting bad recommendations against trustworthy nodes \cite{SecureSocSimil-112} \cite{SocialIoTTrust2016}. We study several recently proposed social trust-based communications in Section \ref{Detection}.

\subsection{Node Attack Models}
\label{NodeAttackModels}
\noindent Opportunistic communications and interactions among mobile nodes in proximity are vulnerable to different types of attacks (\textit{e.g.}, physical attacks, compromised credentials, and protocol attacks) due to the open architecture of the network, node mobility, and privacy issues. To deal with network attacks, numerous protection and defense mechanisms have been designed to guaranty the requirements of a secure communication, such as authentication, availability, confidentiality, and integrity. Despite various attack and defense mechanisms discussed in the literature, in this paper, we focus on different forms of internal attacks (\textit{i.e.}, the attacks launched by nodes with valid cryptographic credentials) that can disrupt the normal communications between the nodes and the network throughput severely.

\begin{figure}[!t]
\centering
\includegraphics[width=3.3 in]{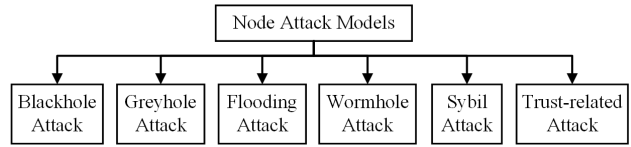}
\caption{Different types of node attacks in wireless relay networks.}
\label{AttackTypes}
\end{figure}

As shown in Fig. \ref{AttackTypes}, we categorize attacks launched by malicious mobile nodes into the following types:

\begin{itemize}
\item \textbf{Blackhole Attack:} a blackhole adversary drops received messages even if it has free buffer space to store them but produces forged metrics (\textit{e.g.}, message delivery probability) to attract more messages or hide its real identity.
\item \textbf{Greyhole Attack:} a particular type of blackhole attack in which a greyhole adversary drops a fraction of received messages even if it has free buffer space but produces forged metrics that makes it difficult for other nodes to detect it. In a complex form of the greyhole attack, the attacker drops some received messages and injects other fake messages instead.
\item \textbf{Data Flooding Attack:} a data flooding attacker injects as many messages as possible into the network to overuse the network resources (\textit{e.g.}, bandwidth, energy, and buffer) and degrade the throughput. A flooding attacker can attack the network in different ways. For example, it may generate fake messages or copy the same message destined for random or selective target nodes through some victim relay nodes that have the highest popularity. In certain cases, a flooding attacker may destine its fake messages to non-existing nodes in order to make them remain in the network longer.
 \item \textbf{Wormhole Attack:} a wormhole adversary receives messages in one location of the network and then moves and replicates them to nodes in another part of the network in order to pretend that messages are transferred through fewer transmission hops. The main objective of a wormhole attacker is to disarrange the topology views of the network by providing fake neighboring information and improve its position (\textit{e.g.}, its reputation).
\item \textbf{Sybil Attack:} a Sybil attacker (or Sybil) generates a large number of bogus identities or location information to establish many fake links in the network with the aim of manipulating its reputation or the bad reputation of other nodes \cite{SybilIoT-55}. For example, a Sybil attempts to disseminate spam and advertisements, produce wrong reports, obtain a disproportionately high benefit from the network without sufficient contribution, and steal the other nodes' private information. In some cases, a mobile Sybil may contact other nodes to share the same social or location information with different forged identities and mislead their routing decisions. Dealing with the Sybil attack becomes more challenging when compromised colluding nodes augment the capability of Sybils.
\item \textbf{Social Trust-related Attacks:} a malicious node can attack a trust management mechanism in different ways to disrupt its functionality. For example, it can launch a self-promoting attack to improve its importance and be selected as the service provider or relay node, but then it refuses to provide the service or provide a malfunctioned service. In addition, malicious nodes can launch other types of trust-related attacks (such as bad-mouthing or ballot stuffing attacks \cite{SocialIoTTrust2016}) in the form of recommendations to exaggerate the trust level of their friends or ruin the reputation of unknown strangers or well-behaved nodes. Thus, a robust trust management mechanism should be designed to protect the trust level of nodes against such attacks.
\end{itemize}

\subsection{Common Data Delivery Protocols to Evaluate Human Non-cooperative Behaviors in WRNs}

\noindent
%Unlike routing in wireless mobile ad hoc networks (\textit{e.g.}, \cite{MANETRouting1, MANETRouting2}), there is a lack of end-to-end connectivity between nodes in WRNs, and nodes employ \textit{store-carry-and-forward} fashion to exchange data with each other. Hence,
In general, data forwarding and dissemination protocols in WRNs employ multi-copy replication mechanisms to improve the data delivery probability with the cost of communication overhead. Broadly, multi-copy replication mechanisms can be classified into two major classes: \textit{stateless} and \textit{deterministic}. In the stateless protocols (\textit{e.g.}, Epidemic \cite{Epidemic-8}, Two-hop \cite{Two-hop-9}, spray and wait (SnW) \cite{SprayWait-10}, and backpressure-based routing \cite{BackpressureDTN2010}), mobile nodes make data replication decisions locally without considering the properties of other nodes (\textit{e.g.}, their delivery probability). In contrast, in deterministic protocols, the nodes' contact history (\textit{e.g.}, Prophet \cite{PROPHET-7}) or social features (\textit{e.g.}, Bubble Rap \cite{BUBBLERap-121}, dLife \cite{dLife-122}, and PIS \cite{PIS-123}) are utilized to choose optimal intermediate nodes and improve the data forwarding performance in terms of important metrics, such as data delivery ratio, delay, and communication overhead.

The majority of works we will discuss through the rest of this paper employ the stateless protocols to evaluate the performance and effectiveness of their solutions in non-cooperative WRNs. We believe that it is because implementing the stateless routing protocols is relatively straightforward. Additionally, the impact of nodes' different behaviors on data delivery performance can be well demonstrated using the stateless protocols.

\section{Impact of Node Selfish Behavior on Opportunistic Communications}
\label{Impact}
\noindent Different forms of nodes' selfish behavior can influence the data delivery performance metrics (\textit{e.g.}, data delivery ratio, delay, transmission cost, and resource consumptions) in different ways. For example, the message dropping or non-forwarding actions of selfish nodes in multi-copy routing protocols can increase the delivery delay but improve the delivery overhead. Moreover, selfish nodes can highly degrade the efficiency of data offloading in D2D communications, especially when seed nodes refuse to deliver the content to non-seed nodes via opportunistic communications. In the literature, different models and techniques have been employed to characterize and estimate how routing metrics change in the presence of non-cooperative nodes. We categorize the impact analysis methods into three classes: theoretical, simulation-based, and hybrid methods. In the following, we discuss the main contributions of each work and highlight their major results.

\subsection{Theoretical Methods}

\begin{figure}[!t]
\centering
\includegraphics[width=3 in]{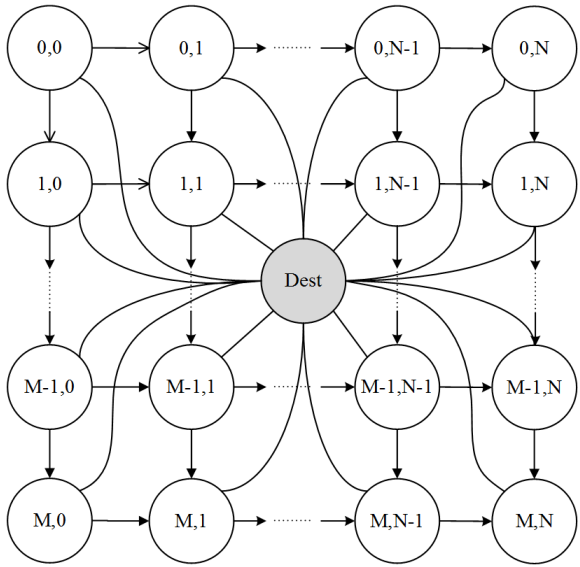}
\caption{A sample two-dimensional CTMC model for the Epidemic routing process with $(M+1)\times(N+1)$ transient states where state (0,0) is the initial state and state (Dest) is the absorbing state \cite{EvaSocSelfish-18}.}
\label{MarkovChain}
\end{figure}

\noindent Several analytical methods have been proposed to analyze the impact of nodes' selfish behavior on the performance of opportunistic communications. A considerable number of theoretical methods in this class have employed the continuous-time Markov chain (CTMC) model to analyze the data delivery process. In general, a CTMC model is characterized by a state space and a transition matrix where the process starts with an initial state and changes to another state according to the probabilities of particular transitions in the transition matrix. Fig. \ref{MarkovChain} shows a CTMC transition machine that models a message relaying in a network with two non-overlapping communities $V_1$, $V_2$ with $N$ and $M$ SS nodes, respectively. The transition process starts from state (0,0) which implies that the number of the message copies in communities $V_1$ and $V_2$ equals to 0. Once the number of the message copies is more than 0, the message may be transmitted to the destination state (Dest). Thus, a major question is how to obtain the transition probability from each state to state (Dest) that can help derive the message delivery performance metrics, such as the delivery delay and cost.

Karaliopoulos \cite{Assessing-11} formulate message relaying in the Epidemic and Two-hop protocols using a two-dimensional CTMC (2D-CTMC) model. In particular, \textit{deceleration factor} metric is devised to measure the deterioration of the delivery delay, which is defined as the ratio of the expected delivery delay when there are \textit{K} selfish nodes versus the case all the nodes are cooperative. The numerical results demonstrate that the delivery delay in both the protocols increases as the number of the selfish nodes goes up. Meanwhile, it is shown that both the protocols are resistant against the selfish behavior when it is probabilistic. For example, the deceleration factor remains below 2 even in the presence of 70\% of selfish nodes with selfishness degree 0.5. Li \textit{et al.} \cite{EvaEff-21} design a 2D-CTMC model to obtain the message delivery delay and cost. The analytical results show that the non-forwarding and non-copying actions have opposite impacts on the Epidemic and Two-hop protocols. For instance, the non-copying action of selfish nodes increases the delivery delay and cost in Two-hop, whereas the delivery delay in Epidemic increases but the cost does not change considerably.

%Similarly, Liu \textit{et al.} \cite{PerfMod-20} design a 2D-CTMC model to study how the cooperation level of selfish nodes influences the performance of Two-hop in terms of data delivery delay and cost. The numerical results reveal that the selfishness degree 0.50 results in a minimum delivery delay but at the expenses of a high delivery cost.

Resta and Santi \cite{FrameworkNonCoopDTN-116} model the routing process in the Epidemic, Two-hop, and SnW protocols as a stochastic coloring process to derive the data delivery delay and communication cost metrics. In particular, three levels of node cooperation: fully cooperative, probabilistic cooperative, and non-cooperative behaviors are considered. Based on the coloring process, a node can be in three states: uncolored (has not received a message), colored active (has at least two copies of the message), and colored inactive (has only one copy of the message that can deliver to its destination). The coloring process finishes when the destination node receives the message and becomes colored. The results show that the data delivery performance doubles even when a small portion of nodes cooperates in message relaying in comparison to the case all the potential forwarders drop messages.

While the above-mentioned studies only consider the nodes' social-oblivious selfishness behavior, Li \textit{et al.} \cite{EvaSocSelfish-18} analyze the impact of SS nodes on the Epidemic routing where the network nodes are partitioned into two non-overlapping communities. In particular, a 2D-CTMC is employed to model the message relaying process. Besides, \textit{delay deceleration ratio} and \textit{cost enhancement ratio} metrics are introduced to measure the performance degradation of the data delivery delay and cost, respectively. The results demonstrate that as the number of selfish nodes increases, the delivery delay increases, but there is more reduction in the delivery cost. Xiao \textit{et al.} \cite{AssessVDTN-24} apply a 2D-CTMC model to explore how IS and SS nodes affect the performance of gossip-based data forwarding in DTNs. The network is partitioned into two non-overlapping communities where the nodes in only one community are IS. The results show that the non-forwarding action of IS nodes reduces the transmission cost more than increasing the delivery delay, whereas the non-copying action of IS nodes degrades the cost less than the delivery delay. Furthermore, the gossip-based forwarding is robust to social selfishness because the transmission cost decreases significantly at the cost of a slight increase in the delivery delay.

\subsection{Simulation-based Methods}

\noindent Several existing studies employ simulations to explore the impact of node selfishness on data delivery performance. The authors in \cite{EffNonCoo-15} explore the impact of the nodes' non-forwarding and partially-forwarding actions on the performance of Epidemic, SnW, and Prophet protocols in terms of the data delivery ratio and delay. The experimental results demonstrate that DTNs tolerate a high percentage of non-cooperative nodes (20-40\% or even 60\%) without too much harm, even though they still utilize the other nodes' resources to deliver their own messages. Meanwhile, synthetic random mobility models are most vulnerable to less cooperation that implies that DTNs are robust against the nodes' non-cooperative behavior. Comparatively, it is revealed that the performance degradation of SnW is relatively higher than Epidemic and Prophet because SnW generates a limited number of message copies.

\begin{figure}[!t]
\centering
\includegraphics[width=3.3 in]{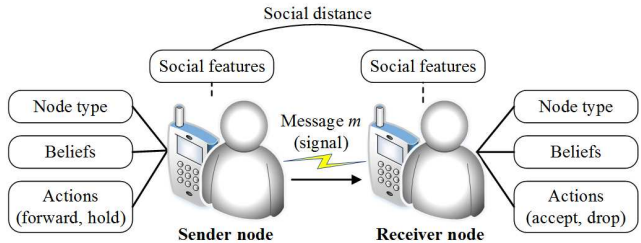}
\caption{The belief-based uncertain interaction between nodes in Sig4UDD.}
\label{Sig4UDDScheme}
\end{figure}

Hui \textit{et al.} \cite{Selfishalturisim-14} study the impact of nodes' different altruistic distributions (such as the percentage of uniform, normal, degree-biased, and community-biased) on the performance of opportunistic communications. The experimental results reveal that a network setting with uniform, normal, or degree-biased distributions can achieve almost 90\% performance of a fully cooperative network due to their multiple forwarding paths. In addition, it is confirmed that the community-biased traffic can further increase the robustness of the network. The authors in \cite{VirtueSelfishness-16} evaluate the performance of a publish-subscribe data offloading system in the presence of the strict and mild nodes. The performance results in terms of the energy consumption and data delivery ratio demonstrate that in the presence of strict nodes, the energy consumption decreases significantly at the cost of losing some data delivery ratio. In contrast, under mild selfishness, the energy consumption further decreases while the delivery ratio increases.

\begin{table*}[!t]
%% increase table row spacing, adjust to taste
%\renewcommand{\arraystretch}{1.3}
% if using array.sty, it might be a good idea to tweak the value of
% \extrarowheight as needed to properly center the text within the cells
\caption{Summary of the works that study the impact of human selfish behavior in wireless relay networks}
\label{tableImpact}
\centering
\linespread{1.1}\selectfont
%% Some packages, such as MDW tools, offer better commands for making tables
%% than the plain LaTeX2e tabular which is used here.
\begin{tabular}{|*{13}{c|}} % repeats {c|} X times
\hline

\multirow{6}*{\rotatebox[origin=c]{90}{\textbf{Approach}}} & 	
%\multirow{6}*{\rotatebox[origin=c]{90}{\textbf{Reference}}} &
\multirow{6}{*}{\parbox[t]{13mm}{\textbf{Reference}}} &	
\multirow{6}*{\parbox[t]{30mm}{\textbf{Principle of proposed solutions}}} &
\multicolumn{2}{c|}{\parbox[t]{11mm}{\textbf{Selfish behavior}}} &
\multicolumn{4}{c|}{\parbox[t]{15mm}{\textbf{Evaluation parameters}}} &
\multicolumn{3}{c|}{\parbox[t]{12mm}{\textbf{Routing protocol}}} &
\multirow{6}{*}{\parbox[t]{45mm}{\textbf{Specialties (+) and limitations (-)}}}
\\

 \cline{4-12}
&&&

\parbox[t]{1mm}{\multirow{4}*{\rotatebox[origin=c]{90}{\textbf{Individual}}}}	&
\parbox[t]{1mm}{\multirow{4}*{\rotatebox[origin=c]{90}{\textbf{Social}}}} &

\parbox[t]{1mm}{\multirow{4}*{\rotatebox[origin=c]{90}{\textbf{Delay}}}} & \parbox[t]{1mm}{\multirow{4}*{\rotatebox[origin=c]{90}{\textbf{Delivery}}}} &
\parbox[t]{1mm}{\multirow{4}*{\rotatebox[origin=c]{90}{\textbf{Cost}}}} &
\parbox[t]{1mm}{\multirow{4}*{\rotatebox[origin=c]{90}{\textbf{Energy}}}} &

\parbox[t]{1mm}{\multirow{4}*{\rotatebox[origin=c]{90}{\textbf{Epidemic}}}}
& \parbox[t]{1mm}{\multirow{4}*{\rotatebox[origin=c]{90}{\textbf{Two-hop}}}}
& \parbox[t]{1mm}{\multirow{4}*{\rotatebox[origin=c]{90}{\textbf{Others}}}} &  \\

\multicolumn{2}{|c|}{}&&&&&&&&&&& \\ \multicolumn{2}{|c|}{}&&&&&&&&&&& \\
\multicolumn{2}{|c|}{}&&&&&&&&&&& \\

\midrule
\hline
%----------------------------------------------------------

\parbox[t]{2mm}{\multirow{18}{0.2in}{\rotatebox[origin=c]{90}{Theoretical}}} &
\multirow{3}{0.7in}{Karaliopoulos \cite{Assessing-11}}&	
\multirow{3}{1.3in}{A 2D-CTMC to model the message relaying process}&	
\multirow{3}*{$\surd$}	&
\multirow{3}*{$\times$}&	
\multirow{3}*{$\surd$}	&
\multirow{3}*{$\times$}&	
\multirow{3}*{$\times$}	&
\multirow{3}*{$\times$}&
\multirow{3}*{$\surd$}	&
\multirow{3}*{$\surd$}	&
\multirow{3}*{$\times$} &
\multirow{3}{1.8in}{+ Considering both non-forwarding and non-copying selfish actions
\\ - Data delivery delay analysis only}  \\ &&&&&&&&&&&& \\ &&&&&&&&&&&& \\  \cline{2-13}
&
\multirow{3}{0.7in}{Li \textit{et al.} \cite{EvaEff-21}}&	
\multirow{3}{1.3in}{A 2D-CTMC to model the message relaying process}&	
\multirow{3}*{$\surd$}	&
\multirow{3}*{$\times$}&	
\multirow{3}*{$\surd$}	&
\multirow{3}*{$\times$}&	
\multirow{3}*{$\surd$}	&
\multirow{3}*{$\times$}&
\multirow{3}*{$\surd$}	&
\multirow{3}*{$\surd$}	&
\multirow{3}*{$\times$} &
\multirow{3}{1.8in}{+ the data delivery delay and cost tradeoff analysis
\\- No comparison with previous works}  \\ &&&&&&&&&&&& \\ &&&&&&&&&&&& \\  \cline{2-13}

%&
%\multirow{3}*{\cite{PerfMod-20}}&	
%\multirow{3}{1.8in}{A complicated 2D-CTMC to model the message delivery}&	
%\multirow{3}*{$\surd$}	&
%\multirow{3}*{$\times$}&	
%\multirow{3}*{$\surd$}	&
%\multirow{3}*{$\times$}&	
%\multirow{3}*{$\surd$}	&
%\multirow{3}*{$\times$}&
%\multirow{3}*{$\times$}	&
%\multirow{3}*{$\surd$}	&
%\multirow{3}*{$\times$} &
%\multirow{3}{1.8in}{+ Analyzing multiple source-destination pairs
%\\- No comparison with previous works} \\ &&&&&&&&&&&& \\ &&&&&&&&&&&& \\  \cline{2-13}

&
\multirow{3}{0.7in}{Resta and \\ Santi \cite{FrameworkNonCoopDTN-116}}&	
\multirow{3}{1.3in}{A stochastic coloring process to model the message delivery}&	
\multirow{3}*{$\surd$}	&
\multirow{3}*{$\times$}&	
\multirow{3}*{$\surd$}	&
\multirow{3}*{$\times$}&	
\multirow{3}*{$\surd$}	&
\multirow{3}*{$\times$}&
\multirow{3}*{$\surd$}	&
\multirow{3}*{$\surd$}	&
\multirow{3}*{$\surd$} &
\multirow{3}{1.8in}{+ Extensive analysis of the effects of selfishness on delivery delay and cost
\\- No comparison with previous work}  \\ &&&&&&&&&&&& \\ &&&&&&&&&&&& \\ \cline{2-13}
&
\multirow{3}{0.7in}{Li \textit{et al.} \cite{EvaSocSelfish-18}}&	
\multirow{3}{1.3in}{A 2D-CTMC to model the message relaying process with social selfishness}&	
\multirow{3}*{$\times$}	&
\multirow{3}*{$\surd$}&	
\multirow{3}*{$\surd$}	&
\multirow{3}*{$\times$}&	
\multirow{3}*{$\surd$}	&
\multirow{3}*{$\times$}&
\multirow{3}*{$\surd$}	&
\multirow{3}*{$\times$}	&
\multirow{3}*{$\times$} &
\multirow{3}{1.8in}{+ Analyzing the impact of selfishness in community-based DTNs
\\- Only evaluates the Epidemic routing}  \\ &&&&&&&&&&&& \\ &&&&&&&&&&&& \\  \cline{2-13}

&
\multirow{3}{0.7in}{Xiao \textit{et al.} \cite{AssessVDTN-24}}&	
\multirow{3}{1.3in}{A 2D-CTMC model to analyze the gossip dissemination}&	
\multirow{3}*{$\surd$}	&
\multirow{3}*{$\surd$}&	
\multirow{3}*{$\surd$}	&
\multirow{3}*{$\times$}&	
\multirow{3}*{$\surd$}	&
\multirow{3}*{$\times$}&
\multirow{3}*{$\surd$}	&
\multirow{3}*{$\times$}	&
\multirow{3}*{$\times$} &
\multirow{3}{1.8in}{+ Studying individual and social selfishness, and system energy
\\- No comparison with previous work}  \\ &&&&&&&&&&&& \\ &&&&&&&&&&&& \\

\hline
%------------------------------------------------------------

\parbox[t]{2mm}{\multirow{27}{0.2in}{\rotatebox[origin=c]{90}{Simulation-based}}}
&

\multirow{3}{0.7in}{Keranen \textit{et al.} \cite{EffNonCoo-15}}&	
\multirow{3}{1.3in}{Evaluating the non-forwarding and partly-forwarding actions}&	
\multirow{3}*{$\surd$}	&
\multirow{3}*{$\times$}&	
\multirow{3}*{$\surd$}	&
\multirow{3}*{$\surd$}&	
\multirow{3}*{$\times$}	&
\multirow{3}*{$\surd$}&
\multirow{3}*{$\surd$}	&
\multirow{3}*{$\surd$}	&
\multirow{3}*{$\surd$} &
\multirow{3}{1.8in}{+ Studying the selfish behavior of a wide class of routing protocols
\\- The selfishness model is weak}  \\ &&&&&&&&&&&& \\ &&&&&&&&&&&& \\ \cline{2-13}
&
\multirow{3}{0.7in}{Hui \textit{et al.} \cite{Selfishalturisim-14}}&	
\multirow{3}{1.3in}{Studying the different distributions of human altruistic models}&	
\multirow{3}*{$\surd$}	&
\multirow{3}*{$\surd$}&	
\multirow{3}*{$\times$}	&
\multirow{3}*{$\surd$}&	
\multirow{3}*{$\times$}	&
\multirow{3}*{$\times$}&
\multirow{3}*{$\surd$}	&
\multirow{3}*{$\times$}	&
\multirow{3}*{$\times$} &
\multirow{3}{1.8in}{+ Analyzing different altruistic behavior and message traffic models
\\- Data delivery ratio analysis only}  \\ &&&&&&&&&&&& \\ &&&&&&&&&&&& \\   \cline{2-13}

&
\multirow{4}{0.7in}{Kouyoumdjieva and Karlsson \cite{VirtueSelfishness-16}}&	
\multirow{4}{1.3in}{Studying the effects of selfishness on publish/subscribe dissemination}&	
\multirow{4}*{$\surd$}	&
\multirow{4}*{$\times$}&	
\multirow{4}*{$\times$}	&
\multirow{4}*{$\surd$}&	
\multirow{4}*{$\times$}	&
\multirow{4}*{$\surd$}&
\multirow{4}*{$\times$}	&
\multirow{4}*{$\times$}	&
\multirow{4}*{$\surd$} &
\multirow{4}{1.8in}{+ Introducing energy-aware selfishness
\\- The lack of design properties}  \\ &&&&&&&&&&&& \\  &&&&&&&&&&&& \\ &&&&&&&&&&&& \\  \cline{2-13}

&
\multirow{3}{0.7in}{Bermejo \textit{et al.} \cite{EmpiricalAlturistic-13}}&	
\multirow{3}{1.3in}{Studying the impact of battery level and social ties on routing performance}&	
\multirow{3}*{$\surd$}	&
\multirow{3}*{$\surd$}&	
\multirow{3}*{$\times$}	&
\multirow{3}*{$\surd$}&	
\multirow{3}*{$\surd$}	&
\multirow{3}*{$\surd$}&
\multirow{3}*{$\surd$}	&
\multirow{3}*{$\times$}	&
\multirow{3}*{$\times$} &
\multirow{3}{1.8in}{+ Applying real-world scenarios and applications in the experiment
\\- Size-dependent performance analysis}  \\ &&&&&&&&&&&& \\ &&&&&&&&&&&& \\ \cline{2-13}

&
\multirow{2}{0.7in}{Xia \textit{et al.} \cite{Sig4UDD-113}}&	
\multirow{2}{1.3in}{A signaling game to analyze the impact of uncertain data forwarding on routing}&	
\multirow{2}*{$\surd$}	&
\multirow{2}*{$\surd$}&	
\multirow{2}*{$\surd$}	&
\multirow{2}*{$\surd$}&	
\multirow{2}*{$\surd$}	&
\multirow{2}*{$\times$}&
\multirow{2}*{$\surd$}	&
\multirow{2}*{$\times$}	&
\multirow{2}*{$\surd$} &
\multirow{2}{1.8in}{+ Introducing a more realistic selfishness model
\\- No consideration of device energy}  \\ &&&&&&&&&&&& \\ &&&&&&&&&&&& \\ \cline{2-13}

&
\multirow{2}{0.7in}{Wang \textit{et al.} \cite{UserUncenrtainty2015}}&	
\multirow{2}{1.3in}{An approximation method based on mean field game to study data diffusion}&	
\multirow{2}*{$\times$}	&
\multirow{2}*{$\surd$}&	
\multirow{2}*{$\surd$}	&
\multirow{2}*{$\times$}&	
\multirow{2}*{$\surd$}	&
\multirow{2}*{$\times$}&
\multirow{2}*{$\surd$}	&
\multirow{2}*{$\times$}	&
\multirow{2}*{$\times$} &
\multirow{2}{1.8in}{+ Comply with a philosophical saying for gossip diffusion in real social life
\\- No analysis of diffusion delay}  \\ &&&&&&&&&&&& \\ &&&&&&&&&&&& \\ \cline{2-13}
&
\multirow{3}{0.7in}{Wang \textit{et al.} \cite{SocialD2DOffload}}&	
\multirow{3}{1.3in}{A network formation game to analyze the opportunistic D2D offloading}&	
\multirow{3}*{$\surd$}	&
\multirow{3}*{$\times$}&	
\multirow{3}*{$\times$}	&
\multirow{3}*{$\surd$}&	
\multirow{3}*{$\times$}	&
\multirow{3}*{$\times$}&
\multirow{3}*{$\surd$}	&
\multirow{3}*{$\times$}	&
\multirow{3}*{$\times$} &
\multirow{3}{1.8in}{+ Introducing the first selfish-aware D2D offloading model
\\- No consideration of SS nodes}  \\ &&&&&&&&&&&& \\ &&&&&&&&&&&& \\ \cline{2-13}
&
\multirow{4}{0.7in}{Wang \textit{et al.} \cite{SSD2DComm2015}}&	
\multirow{4}{1.3in}{A matching solution to analyze SS nodes on community-based D2D communications}&	
\multirow{4}*{$\times$}	&
\multirow{4}*{$\surd$}&	
\multirow{4}*{$\times$}	&
\multirow{4}*{$\surd$}&	
\multirow{4}*{$\times$}	&
\multirow{4}*{$\times$}&
\multirow{4}*{$\surd$}	&
\multirow{4}*{$\times$}	&
\multirow{4}*{$\times$} &
\multirow{4}{1.8in}{+ The first work to analyze SS nodes in D2D communications
\\- No evaluation of other metrics}  \\ &&&&&&&&&&&& \\ &&&&&&&&&&&& \\ &&&&&&&&&&&& \\ \cline{2-13}
&
\multirow{3}{0.7in}{Cao \textit{et al.} \cite{SelfishD2DYongLi}}&	
\multirow{3}{1.3in}{Analyzing node selfishness in the BS-to-device and D2D communications}&	
\multirow{3}*{$\surd$}	&
\multirow{3}*{$\surd$}&	
\multirow{3}*{$\times$}	&
\multirow{3}*{$\surd$}&	
\multirow{3}*{$\times$}	&
\multirow{3}*{$\times$}&
\multirow{3}*{$\surd$}	&
\multirow{3}*{$\times$}	&
\multirow{3}*{$\times$} &
\multirow{3}{1.8in}{+ Considering both IS and SS nodes
\\- Requires all the link information to establish the time-varying graph}  \\ &&&&&&&&&&&& \\ &&&&&&&&&&&& \\
\hline
%----------------------------------------------

\parbox[t]{2mm}{\multirow{14}{0.2in}{\rotatebox[origin=c]{90}{Hybrid}}}
&

\multirow{3}{0.7in}{Ip \textit{et al.} \cite{PerfModEpid-23}}&	
\multirow{3}{1.3in}{An ODE model to analyze probabilistic selfish behavior}&	
\multirow{3}*{$\surd$}	&
\multirow{3}*{$\times$}&	
\multirow{3}*{$\surd$}	&
\multirow{3}*{$\surd$}&	
\multirow{3}*{$\surd$}	&
\multirow{3}*{$\times$}&
\multirow{3}*{$\surd$}	&
\multirow{3}*{$\surd$}	&
\multirow{3}*{$\times$} &
\multirow{3}{1.8in}{+ Considering nodes with different transmission ranges
\\- No analysis of social properties} \\ &&&&&&&&&&&& \\ &&&&&&&&&&&& \\  \cline{2-13}

&
\multirow{2}{0.7in}{Li \textit{et al.} \cite{ImpactMulti-26}}&	
\multirow{2}{1.3in}{A 3D-CTMC to model the message multicasting}&	
\multirow{2}*{$\surd$}	&
\multirow{2}*{$\surd$}&	
\multirow{2}*{$\surd$}	&
\multirow{2}*{$\times$}&	
\multirow{2}*{$\surd$}	&
\multirow{2}*{$\times$}&
\multirow{2}*{$\surd$}	&
\multirow{2}*{$\surd$}	&
\multirow{2}*{$\times$} &
\multirow{2}{1.8in}{+ Considers both IS and SS nodes
\\- The network model is not general}  \\ &&&&&&&&&&&& \\ \cline{2-13}
&
\multirow{3}{0.7in}{Wu \textit{et al.} \cite{ModLimEpi-22}}&	
\multirow{3}{1.3in}{An ODE model to study the impact of IS and SS nodes on routing}&	
\multirow{3}*{$\surd$}	&
\multirow{3}*{$\surd$}&	
\multirow{3}*{$\times$}	&
\multirow{3}*{$\surd$}&	
\multirow{3}*{$\times$}	&
\multirow{3}*{$\surd$}&
\multirow{3}*{$\surd$}	&
\multirow{3}*{$\surd$}	&
\multirow{3}*{$\times$} &
\multirow{3}{1.8in}{+ Validating the theoretical results using simulations
\\-Unrealistic routing assumptions}  \\ &&&&&&&&&&&& \\ &&&&&&&&&&&& \\ \cline{2-13}

&
\multirow{3}{0.7in}{Sermpezis and Spyropoulos \cite{UndSelfish-27}}&	
\multirow{3}{1.3in}{A generic model to analysis the influence of SS nodes on routing based on mobility}&	
\multirow{3}*{$\times$}	&
\multirow{3}*{$\surd$}&	
\multirow{3}*{$\surd$}	&
\multirow{3}*{$\surd$}&	
\multirow{3}*{$\surd$}	&
\multirow{3}*{$\surd$}&
\multirow{3}*{$\surd$}	&
\multirow{3}*{$\surd$}	&
\multirow{3}*{$\surd$} &
\multirow{3}{1.8in}{+ Analyzing social selfishness policies
\\- The nodes' social ties are identified based on only contact history} \\ &&&&&&&&&&&& \\ &&&&&&&&&&&& \\ \cline{2-13}
&
\multirow{4}{0.7in}{Sermpezis and Spyropoulos \cite{DelayAnl-117}}&	
\multirow{4}{1.3in}{An asymptotic model to analyze the impact of SS nodes on the performance of stateless routing}&	
\multirow{4}*{$\times$}	&
\multirow{4}*{$\surd$}&	
\multirow{4}*{$\surd$}	&
\multirow{4}*{$\times$}&	
\multirow{4}*{$\times$}	&
\multirow{4}*{$\times$}&
\multirow{4}*{$\surd$}	&
\multirow{4}*{$\surd$}	&
\multirow{4}*{$\surd$} &
\multirow{4}{1.8in}{+ Analyzing delivery delay in heterogeneous networking scenarios
\\- Data delivery delay analysis only}  \\ &&&&&&&&&&&& \\ &&&&&&&&&&&& \\ &&&&&&&&&&&& \\
\hline
\multicolumn{13}{c}{(``$\surd$'' if the protocol satisfies the property, ``$\times$'' if not)}
\end{tabular}
\end{table*}

While the studies above focus on the nodes' social-oblivious selfishness behavior, Bermejo \textit{et al.} \cite{EmpiricalAlturistic-13} study human altruism in AppExp and WebExp applications with respectively 800 and 737 nodes considering the nodes' remaining battery level and social tie information. The experiments show that nodes respectively exhibit 70\% and 52\% altruistic behavior in AppExp and WebExp when a minor credit (1 dollar) is awarded. Meanwhile, the nodes are not willing to relay data received from others when their remaining battery level is less than 10\%. Xia \textit{et al.} \cite{Sig4UDD-113} explore the impact of IS and SS nodes on social-based routing protocols under uncertain node cooperation. In particular, a signaling game approach (Sig4UDD) is proposed where Bayesian Nash equilibrium and perfect Bayesian equilibrium are employed to analyze the nodes' one-stage and multi-stage interactions (Fig. \ref{Sig4UDDScheme}). Meanwhile, a belief system is established to help the nodes predict the type of their encounters and decide whether to forward a message to them or not. The experimental results demonstrate that nodes in Sig4UDD can effectively establish their beliefs based on their previous interactions that can decrease the transmission cost significantly while improving the data delivery delay. Similar to \cite{Sig4UDD-113}, Wang \textit{et al.} \cite{UserUncenrtainty2015} employ random utility theory to model gossip diffusion of rational nodes in social networks under uncertainty. Next, a formal framework based on mean field theory is devised to analyze the diffusion process. The results demonstrate that small uncertainty can speed up gossip diffusing significantly.

A limited number of proposals study the impact of human selfish behavior on the performance of D2D communications. The authors in \cite{SocialD2DOffload} consider an opportunistic data offloading approach in network-assisted D2D communications in which nodes download the contents from the BS and then decide whether to share them with other nodes or not according to their historical records. Next, a network formation game is designed to model the dynamic characteristics of the nodes' selfish behaviors wherein the gain and cost functions are specified for downloading content via D2D communications or cellular network. The simulation results show that the selfish behavior of nodes can degrade the offloading efficiency significantly. In addition, the cost ratio between the cellular and D2D transmissions, as well as the nodes' access delays and mobility patterns affect the performance gap significantly.

\begin{figure}[!t]
\centering
\includegraphics[width=3.4 in]{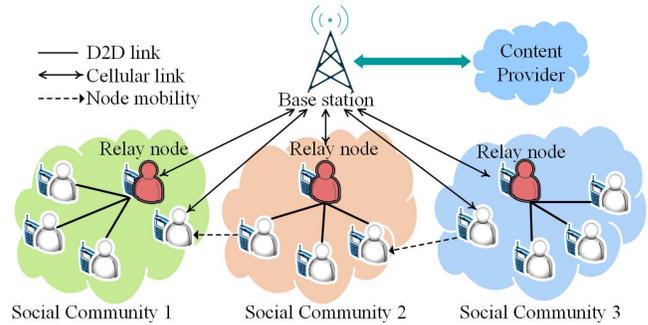}
\caption{Community-based D2D communications underlying cellular network in the presence of socially selfish relay nodes.}
\label{D2DOffloading}
\end{figure}

Wang \cite{SSD2DComm2015} study human selfish behavior in community-based D2D communications in which SS nodes in each community participate in relaying contents received from the BS to non-relay nodes with respect to their social relations, as shown in Fig. \ref{D2DOffloading}. The study adopts a bipartite graph to obtain a matching solution between the relay and non-relay nodes when the cooperation degree of relay nodes and the number of communities vary. The experiments show that SS nodes degrade the system throughput with fewer mobile devices. Besides, it is revealed that the highest performance gap occurs when the number of relay and non-relay nodes are equal. Similarly, Gao \textit{et al.} \cite{SelfishD2DYongLi} employ a time-varying graph model to study the impact of IS and SS nodes on the performance of data offloading in community-based D2D communications. It is assumed that a BS transmits data to a helper seed node and requests it to disseminate the data to the subscribers or other seed nodes. Nevertheless, a selfish seed node can exhibit selfish behavior in receiving contents from the BS or forwarding them to subscribers. The experimental results demonstrate that a few numbers of IS and SS nodes inside each community do not affect the network throughput considerably, especially in a network with a large number of communities.

\subsection{Hybrid Methods}

\noindent The majority of studies in this class employ theoretical approaches to model and analyze opportunistic data delivery process and then conduct simulations to validate the theoretical results. Manam \textit{et al.} \cite{PerfModEpid-23} apply ordinary differential equation (ODE) model to analyze the impact of selfish nodes with probabilistic non-forwarding and non-copying actions and egotistic nodes (\textit{i.e.}, nodes with different communication ranges) on the performance of the Epidemic and Two-hop protocols. The numerical and simulation-based results in the presence of 50\% of selfish and 50\% egotistic nodes show that the delivery ratio goes up, the delay decreases, and the cost increases as the number of nodes increases from 0 to 70.

Unlike \cite{PerfModEpid-23} that only addresses IS nodes, Li \textit{et al.} \cite{ImpactMulti-26} employ a 3D-CTMC model to evaluate the impact of the IS and SS nodes on the performance of Two-hop multicast in DTNs. To model the social selfishness, the network is divided into three non-overlapping communities $V_1$, $V_2$, and $V_3$ , based on which the source and multicast destination nodes are placed in $V_1$ and $V_3$ and the IS nodes are placed in $V_2$. The numerical results show that the data delivery delay increases as the number of IS nodes increases. Additionally, it is concluded that the non-copying action of SS nodes affects the data delivery performance considerably. Wu \textit{et al.} \cite{ModLimEpi-22} apply the ODE model to evaluate the influence of IS and SS nodes on the performance of community-based DTNs using the Epidemic and Two-hop protocols. It is assumed that the network is divided into multiple communities where IS nodes do not relay messages to other nodes in the same community, whereas SS nodes relay messages to nodes in the same community. The experimental results demonstrate that the data delivery ratio decreases as the number of communities increases.

The authors in \cite{UndSelfish-27} study the impact of SS nodes on opportunistic data delivery performance by modeling different cooperation policies where the cooperation level of SS nodes is identified based on their contact rates. First, closed-form expressions are derived to approximate the expected data delivery delay with respect to a broad range of mobility scenarios. Next, simulations are conducted to validate the theoretical results using the synthetic and realistic mobility traces. The numerical results demonstrate that complex selfishness policies cannot achieve better performance than a uniform policy for power versus delay tradeoffs, whereas they can optimize power versus delivery ratio tradeoffs. The same authors in \cite{DelayAnl-117} investigate the impact of SS nodes on the delivery delay in the Epidemic, Two-hop, and SnW protocols with heterogeneous contact distributions. The analytical expressions prove that a first-order mean value approximation for the basic epidemic spreading step becomes exact in large-scale networks.

\textbf{Summary:} Table \ref{tableImpact} summarizes the important features of the research we studied in this section. It can be seen that a limited number of analytical techniques and tools (such as CTMC and ODE models) are employed to study the impact of human selfish behavior on the performance of data delivery protocols in WRNs, in comparison to other fields, such as opportunistic scheduling in opportunistic communication \cite{OppScheduling}. Meanwhile, almost all of the existing studies explore the data delivery delay and transmission cost parameters and do not study the other important parameters, such as the delivery ratio and energy consumption. In addition, there is a lack of an analytical technique to quantify the impact of human selfish behavior on D2D communications. Furthermore, despite the fact that a considerable number of simulation-based experiments study the human non-cooperative behavior in DTNs, the impact of human behavior in terms of different parameters (\textit{e.g.}, delivery delay, transmission cost, and energy consumption) are not explored in D2D communications sufficiently.

\section{Selfish and Malicious Node Detection and Isolation Mechanisms}
\label{Detection}
\noindent Detecting non-cooperative nodes and disseminating the detection information through the network can reduce the loss of network resources. Nevertheless, designing an effective detection and defense system in WRNs is extremely challenging due to the intermittent node connectivity and dynamic network topology. In other words, the misbehaving actions of selfish and malicious nodes are spread in space and time, and the observations of one node might not sufficiently indicate the misbehavior of its encountered nodes. This issue becomes more challenging when attackers collude with each other to boost their metrics and deceive the detection system. In the rest of this section, we discuss well-known selfish and misbehavior detection and defense systems in WRNs.

\noindent
\subsection{Selfish Node Detection Systems}
We categorize selfish node detection schemes into two classes: watchdog systems and social trust-based communications. In the rest of this subsection, we introduce well-known works in each category and discuss their properties.

\begin{figure}[!t]
\centering
\includegraphics[width=3.4 in]{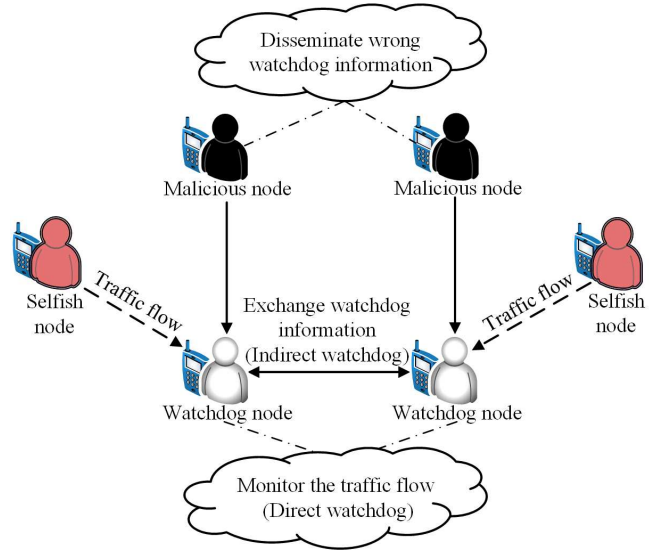}
\caption{A general watchdog scenario in non-cooperative WRNs.}
\label{WatchdogOverview}
\end{figure}

\subsubsection{Watchdog Systems}
as shown in Fig. \ref{WatchdogOverview}, trusted watchdog nodes in watchdog systems analyze the traffic received from their encountered nodes to decide whether they have selfish behavior in message relaying or not (direct watchdog). However, inter-contact times (\textit{i.e.}, two consecutive contacts) between nodes in WRNs can be quite long. Hence, the watchdog nodes may not receive sufficient direct watchdog information to judge the behavior of other nodes. Thus, they can share their opinions about other nodes with each other that help them detect the selfish nodes swiftly and accurately (indirect watchdog). When a node is detected as a non-cooperative node, it is called a \textit{positive detection} (or positive); otherwise, it is called \textit{negative detection} (or negative). However, due mainly to the wrong watchdog information disseminated by malicious nodes, a watchdog node may detect a cooperative node as non-cooperative (false positive) or a non-cooperative node as cooperative (false negative) that can degrade the performance of the watchdog system severely.

Although several watchdog systems have been designed for wireless ad hoc networks (\textit{e.g.}, \cite{AckMisbehv-61,GreedyWatchdog-62}), they cannot be applied to WRNs due to their unique characteristics. A major reason is that the sender of a message in ad hoc networks can observe the relaying behavior of nodes in the delivery path of the message due to the end-to-end node connectivity. Thus, the sender node can detect the nodes' selfish behavior by analyzing the traffic on the message delivery path. In contrast, the observation of a node in WRNs may not indicate the selfish behavior of other nodes due to the intermittent node connectivity. Therefore, the node has to investigate the consistency of the history of contact and message exchange records (observed directly or received from other nodes) to detect selfish message droppers.

Recently, a number of cooperative watchdog systems have been proposed in DTNs. The authors in \cite{EvalSelfiDetect-63} propose a contact history-based collaborative watchdog scheme in which the watchdog nodes use both the direct and indirect watchdog information to detect selfish nodes.
To reduce the impact of false positives and negatives, a controlled mixed diffusion method is applied where the positive detections are always diffused but a fraction of the negative detections is disseminated. Additionally, a 2D-CTMC model is designed to evaluate the detection time and ratio. The experiments show that the proposed scheme reduces the detection time from 20\% for a very low degree of collaboration to 99\% for higher degrees of collaboration. The extension of \cite{EvalSelfiDetect-63} is CoCoWa \cite{CoCoWa-64} in which a reputation scheme is designed to protect the watchdog system against the wrong watchdog information generated by malicious nodes.

Ayday and Fekri \cite{IterDetec-67} propose a graph-based iterative algorithm, namely ITRM, to detect and isolate message droppers in DTNs. In ITRM, watchdog nodes store a rating table about the reputation of other nodes, which is updated based on their direct and indirect watchdog information. The rating table is represented by a bipartite graph where a check vertex shows a watchdog node, and bit vertices show all the nodes that the watchdog node has received watchdog information from them. When two nodes contact each other, they exchange a receipt for each received message along with a signed timestamp, based on which the watchdog nodes can detect message droppers. However, ITRM uses a binary reputation where the reputation (\textit{i.e.}, type) of nodes can change easily if contradictory watchdog information is received.
Similarly, Dias \textit{et al.} \cite{CoopWatch-65} propose a reputation-based cooperative watchdog system to detect message droppers in which the reputation of nodes is updated based on their relayed and delivered messages. In addition, encountered nodes share their opinions about other nodes with each other to improve the detection performance. Finally, selfish and cooperative nodes are punished or rewarded, respectively.

Zhu \textit{et al.} \cite{ProbDetect-68} propose a probabilistic detection scheme, namely iTrust, where a trusted authority (TA) checks the behavior of nodes based on their forwarding history evidence. To achieve a trade-off between the detection accuracy and cost, a reputation system is designed in iTrust where nodes with a good reputation are checked with a low frequency while suspicious nodes are checked with a high frequency. Moreover, an inspection game is played between an inspector (\textit{i.e.}, TA) and an inspectee to find an optimal investigation probability and ensure that message droppers can be detected with a high accuracy and low communication overhead.

\subsubsection{Social Trust-based Systems}
\label{SocTrustManag}
establishing social trust relationships between mobile nodes by leveraging their online social information (explicit trust) as well as their interactions or mobility properties (implicit trust) can help select trusted and secured relay nodes, thus improve the data delivery in WRNs \cite{SocTrustOppNets}. In other words, social trust-based data relaying can avoid selfish nodes, thus stimulating them to cooperative in data forwarding \cite{TrustSocIoT2014}. In addition, it can protect the network against social trust-related malicious attacks, which will be discussed in Subsection \ref{SocialTrustAttack}. However, establishing trust relations and propagating them in infrastructure-less wireless networks are very challenging because there is no centralized authority. In this subsection, we introduce well-known social trust management mechanisms in non-cooperative WRNs and discuss their properties.

\begin{table*}[!t]
%% increase table row spacing, adjust to taste
%\renewcommand{\arraystretch}{1.3}
% if using array.sty, it might be a good idea to tweak the value of
% \extrarowheight as needed to properly center the text within the cells
\caption{Summary of the selfish node detection and isolation mechanisms in wireless relay networks}
\label{tableWatchdog}
\centering
%% Some packages, such as MDW tools, offer better commands for making tables
%% than the plain LaTeX2e tabular which is used here.
\begin{tabular}{|*{11}{c|}} % repeats {c|} X times
\hline

\multirow{8}{*}{\rotatebox[origin=c]{90}{\textbf{Detection method}}}  & 	
%\multirow{8}{*}{\rotatebox[origin=c]{90}{\textbf{Reference}}} &	
\multirow{8}{*}{\parbox[t]{13mm}{\textbf{Reference}}} &
\multirow{8}{*}{\parbox[t]{40mm}{\textbf{Principle of proposed solutions}}} &
\multicolumn{3}{c|}{\parbox[t]{10mm}{\textbf{Influence analyze}}} &
\multicolumn{3}{c|}{\parbox[t]{16mm}{\textbf{Protection information}}} &
\multirow{8}{*}{\rotatebox[origin=c]{90}{\textbf{Fully distributed}}} &	
\multirow{8}{*}{\parbox[t]{45mm}{\textbf{Specialties (+) and limitations (-)}}}  \\
\cline{4-9}
&&&
\parbox[t]{1mm}{\multirow{7}*{\rotatebox[origin=c]{90}{\textbf{Misreport}}}}	&
\parbox[t]{1mm}{\multirow{7}*{\rotatebox[origin=c]{90}{\textbf{False positive}}}}	&
\parbox[t]{1mm}{\multirow{7}*{\rotatebox[origin=c]{90}{\textbf{False negative}}}} &

\parbox[t]{1mm}{\multirow{7}*{\rotatebox[origin=c]{90}{\textbf{Contact records}}}}  & \parbox[t]{1mm}{\multirow{7}*{\rotatebox[origin=c]{90}{\textbf{Message records}}}} &
\parbox[t]{1mm}{\multirow{7}*{\rotatebox[origin=c]{90}{\textbf{Social features}}}} && \\

\multicolumn{2}{|c|}{}&&&&&&&&& \\ \multicolumn{2}{|c|}{}&&&&&&&&& \\
\multicolumn{2}{|c|}{}&&&&&&&&& \\ \multicolumn{2}{|c|}{}&&&&&&&&& \\  \multicolumn{2}{|c|}{}&&&&&&&&& \\ \multicolumn{2}{|c|}{}&&&&&&&&& \\

\midrule
\hline

\parbox[t]{2mm}{\multirow{15}{0.2in}{\rotatebox[origin=c]{90}{Watchdog Systems}}} &
\multirow{3}{0.7in}{Hern\'{a}ndez-Orallo \textit{et al.} \cite{EvalSelfiDetect-63}}&	
\multirow{3}{1.7in}{A 2D-CTMC model to evaluate the selfish node detection time and overhead}&	
\multirow{3}*{$\times$}	&
\multirow{3}*{$\surd$}&	
\multirow{3}*{$\surd$}	&
\multirow{3}*{$\surd$}&	
\multirow{3}*{$\surd$}	&
\multirow{3}*{$\times$}&
\multirow{3}*{$\surd$}	&
\multirow{3}{2in}{+ Evaluation of the effects of false positives and negatives on detection performance
\\- No consideration of malicious behavior}  \\ &&&&&&&&&& \\ &&&&&&&&&& \\ \cline{2-11}
&
\multirow{3}{0.7in}{Hern\'{a}ndez-Orallo \textit{et al.} \cite{CoCoWa-64}}&	
\multirow{3}{1.7in}{A 4D-CTMC model to detect selfish nodes and cope with malicious nodes}&	
\multirow{3}*{$\surd$}	&
\multirow{3}*{$\surd$}&	
\multirow{3}*{$\surd$}	&
\multirow{3}*{$\surd$}&	
\multirow{3}*{$\surd$}	&
\multirow{3}*{$\times$}&
\multirow{3}*{$\surd$}	&
\multirow{3}{2in}{+ Combines the collaboration with reputation
\\- No consideration of social behavior} \\ &&&&&&&&&&  \\ &&&&&&&&&& \\ \cline{2-11}

&
\multirow{3}{0.7in}{Ayday and Fekri \cite{IterDetec-67}}&	
\multirow{3}{1.7in}{A graph-based iterative algorithm to detect malicious nodes}&	
\multirow{3}*{$\surd$}	&
\multirow{3}*{$\times$}&	
\multirow{3}*{$\times$}	&
\multirow{3}*{$\surd$}&	
\multirow{3}*{$\times$}	&
\multirow{3}*{$\times$}&
\multirow{3}*{$\surd$}	&
\multirow{3}{2in}{+ Combines QoS trust and reputation
\\- No evaluation of the nodes' selfish behavior} \\ &&&&&&&&&& \\ &&&&&&&&&&  \\ \cline{2-11}
&
\multirow{3}{0.7in}{Dias \textit{et al.} \cite{CoopWatch-65}}&	
\multirow{3}{1.7in}{A cooperative selfish node detection mechanism based on node reputation}&	
\multirow{3}*{$\times$}	&
\multirow{3}*{$\times$}&	
\multirow{3}*{$\times$}	&
\multirow{3}*{$\surd$}&	
\multirow{3}*{$\times$}	&
\multirow{3}*{$\times$}&
\multirow{3}*{$\surd$}	&
\multirow{3}{2in}{+ Realistic evaluation scenarios
\\- No evaluation of false positives and negatives}  \\ &&&&&&&&&&  \\ &&&&&&&&&&  \\ \cline{2-11}
&
\multirow{3}{0.7in}{Zhu \textit{et al.} \cite{ProbDetect-68}}&	
\multirow{3}{1.7in}{A probabilistic misbehavior detection scheme based on inspection game}&	
\multirow{3}*{$\surd$}	&
\multirow{3}*{$\surd$}&	
\multirow{3}*{$\surd$}	&
\multirow{3}*{$\surd$}&	
\multirow{3}*{$\surd$}	&
\multirow{3}*{$\times$}&
\multirow{3}*{$\times$}	&
\multirow{3}{2in}{+ Achieves a high detection ratio with low communication overhead
\\- Depends on a centralized third party}  \\ &&&&&&&&&& \\ &&&&&&&&&& \\
\hline
\parbox[t]{2mm}{\multirow{30}{0.2in}{\rotatebox[origin=c]{90}{Social Trust-based Systems}}} &
\multirow{3}{0.7in}{Bigwood and Henderson \cite{IRONMAN-77}}&	
\multirow{3}{1.7in}{A trust mechanism based on the self-reported social networks to detect selfish nodes}&	
\multirow{3}*{$\times$}	&
\multirow{3}*{$\surd$}&	
\multirow{3}*{$\times$}	&
\multirow{3}*{$\surd$}&	
\multirow{3}*{$\surd$}	&
\multirow{3}*{$\surd$}&
\multirow{3}*{$\surd$}	&
\multirow{3}{2in}{+ A simple benchmark detection method
\\- Selfish nodes have chance to only forward their own messages before being detected} \\ &&&&&&&&&& \\ &&&&&&&&&&  \\ \cline{2-11}
&
\multirow{3}{0.7in}{Ciobanu \textit{et al.} \cite{SENSE-66}}&	
\multirow{3}{1.7in}{A social and content-based selfish node detection scheme}&	
\multirow{3}*{$\times$}	&
\multirow{3}*{$\times$}&	
\multirow{3}*{$\times$}	&
\multirow{3}*{$\surd$}&	
\multirow{3}*{$\times$}	&
\multirow{3}*{$\surd$}&
\multirow{3}*{$\surd$}	&
\multirow{3}{2in}{+ Considers both individual and social aspects of human altruism
\\- Considers a binary social tie relation} \\ &&&&&&&&&& \\ &&&&&&&&&&  \\ \cline{2-11}
&
\multirow{4}{0.7in}{Chen \textit{et al.} \cite{DynTrustRouting2014}}&	
\multirow{4}{1.7in}{A social trust management scheme to minimize trust bias and maximize the routing performance}&	
\multirow{4}*{$\surd$}	&
\multirow{4}*{$\surd$}&	
\multirow{4}*{$\surd$}	&
\multirow{4}*{$\surd$}&	
\multirow{4}*{$\surd$}	&
\multirow{4}*{$\surd$}&
\multirow{4}*{$\surd$}	&
\multirow{4}{2in}{+ Deals with both selfish behavior and trust-related attacks
\\- Lack of analytical evaluations} \\ &&&&&&&&&& \\ &&&&&&&&&&  \\ &&&&&&&&&&  \\ \cline{2-11}
&
\multirow{4}{0.7in}{Yao \textit{et al.} \cite{SecureSocSimil-112}}&	
\multirow{4}{1.7in}{A trust mechanism based on the social similarity to select trustworthy relay nodes}&	
\multirow{4}*{$\surd$}	&
\multirow{4}*{$\times$}&	
\multirow{4}*{$\times$}	&
\multirow{4}*{$\surd$}&	
\multirow{4}*{$\times$}	&
\multirow{4}*{$\surd$}&
\multirow{4}*{$\surd$}	&
\multirow{4}{2in}{+ Exploits nodes' contact history and social features to identify their trust relationships
\\- No evaluation of false positives and negatives} \\ &&&&&&&&&& \\ &&&&&&&&&& \\ &&&&&&&&&& \\

\cline{2-11}
&
\multirow{3}{0.7in}{Chen \textit{et al.} \cite{SocialIoTTrust2016}}&	
\multirow{3}{1.7in}{An adaptive trust management mechanism for social IoT systems}&	
\multirow{3}*{$\surd$}	&
\multirow{3}*{$\times$}&	
\multirow{3}*{$\times$}	&
\multirow{3}*{$\surd$}&	
\multirow{3}*{$\times$}	&
\multirow{3}*{$\surd$}&
\multirow{3}*{$\surd$}	&
\multirow{3}{2in}{+ Tunes the best trust parameters in response to changing the system conditions
\\- Lack of analytical evaluations} \\ &&&&&&&&&& \\ &&&&&&&&&&  \\

\cline{2-11}
&
\multirow{3}{0.7in}{Ometov \textit{et al.} \cite{TrustedSocD2DOmetov}}&	
\multirow{3}{1.7in}{A coalitional game approach to cluster nodes based on their trust level}&	
\multirow{3}*{$\times$}	&
\multirow{3}*{$\times$}&	
\multirow{3}*{$\times$}	&
\multirow{3}*{$\times$}&	
\multirow{3}*{$\times$}	&
\multirow{3}*{$\surd$}&
\multirow{3}*{$\times$}	&
\multirow{3}{2in}{+ Discussing several possible future research directions
\\- No consideration of trust-related attacks} \\ &&&&&&&&&& \\ &&&&&&&&&&  \\
\cline{2-11}
&
\multirow{3}{0.7in}{Chen \textit{et al.} \cite{ExploitSocTieD2D2015}}&	
\multirow{3}{1.7in}{A coalitional game to establish trusted D2D communications based on social ties}&	
\multirow{3}*{$\times$}	&
\multirow{3}*{$\times$}&	
\multirow{3}*{$\times$}	&
\multirow{3}*{$\surd$}&	
\multirow{3}*{$\times$}	&
\multirow{3}*{$\surd$}&
\multirow{3}*{$\times$}	&
\multirow{3}{2in}{+ Considers both social trust and social reciprocity in relay selection
\\- Only considers in-band communications} \\ &&&&&&&&&& \\ &&&&&&&&&&  \\

\cline{2-11}
&
\multirow{4}{0.7in}{Militano \textit{et al.} \cite{TrustIoT2016}}&	
\multirow{4}{1.7in}{A coalitional game for multi-hop content offloading in network-assisted D2D communications}&	
\multirow{4}*{$\times$}	&
\multirow{4}*{$\times$}&	
\multirow{4}*{$\times$}	&
\multirow{4}*{$\surd$}&	
\multirow{4}*{$\times$}	&
\multirow{4}*{$\surd$}&
\multirow{4}*{$\times$}	&
\multirow{4}{2in}{+ The combination of social relationships and reputation to identify nodes' trust level
\\- No consideration of trust-related attacks} \\ &&&&&&&&&& \\ &&&&&&&&&&  \\ &&&&&&&&&&  \\

%\cline{2-11}
%&
%\multirow{3}*{Ometov \textit{et al.} \cite{SicTrustD2DImplement}}&	
%\multirow{3}{2in}{A proof-of-concept implementation of social-aware D2D communications}&	
%\multirow{3}*{$\times$}	&
%\multirow{3}*{$\times$}&	
%\multirow{3}*{$\times$}	&
%\multirow{3}*{$\times$}&	
%\multirow{3}*{$\times$}	&
%\multirow{3}*{$\surd$}&
%\multirow{3}*{$\times$}	&
%\multirow{3}{2.2in}{+ Evaluation of a real D2D messaging scenario
%} \\ &&&&&&&&&& \\ &&&&&&&&&&  \\

\cline{2-11}
&
\multirow{4}{0.7in}{Zhang \textit{et al.} \cite{StoppingGameD2D}}&	
\multirow{4}{1.7in}{A stoping theory to choose trusted relay nodes in D2D communications based on their physical and social information}&	\multirow{4}*{$\times$}	&
\multirow{4}*{$\times$}&	
\multirow{4}*{$\times$}	&
\multirow{4}*{$\surd$}&	
\multirow{4}*{$\times$}	&
\multirow{4}*{$\surd$}&
\multirow{4}*{$\times$}	&
\multirow{4}{2in}{+ An effective model to update nodes' reputation and detect their selfishness
\\- Privacy concerns because of revealing the nodes' location information  } \\ &&&&&&&&&& \\ &&&&&&&&&&  \\ &&&&&&&&&&  \\

\cline{2-11}
&
\multirow{4}{0.7in}{Yan \textit{et al.} \cite{PartnerSelection2017}}&	
\multirow{4}{1.7in}{A rough set algorithm to select trustworthy relay nodes based on multi-dimensional trust relationships}&	
\multirow{4}*{$\times$}	&
\multirow{4}*{$\times$}&	
\multirow{4}*{$\times$}	&
\multirow{4}*{$\surd$}&	
\multirow{4}*{$\surd$}	&
\multirow{4}*{$\surd$}&
\multirow{4}*{$\times$}	&
\multirow{4}{2in}{+ The psychological structure of users are considered
\\- No consideration of trust-related attacks} \\ &&&&&&&&&& \\ &&&&&&&&&&  \\ &&&&&&&&&&  \\

\cline{2-11}
&
\multirow{4}{0.7in}{Cao \textit{et al.} \cite{SocialTrsutCoalition2015}}&	
\multirow{4}{1.7in}{A group-based video multicast system based on social trust and reciprocity in D2D communications}&	
\multirow{4}*{$\times$}	&
\multirow{4}*{$\times$}&	
\multirow{4}*{$\times$}	&
\multirow{4}*{$\times$}&	
\multirow{4}*{$\times$}	&
\multirow{4}*{$\surd$}&
\multirow{4}*{$\surd$}	&
\multirow{4}{2in}{+ Employing real-world video traces} \\ &&&&&&&&&& \\ &&&&&&&&&&  \\ &&&&&&&&&&  \\

\hline
\multicolumn{11}{c}{(``$\surd$'' if the protocol satisfies the property, ``$\times$'' if not)}
%\multicolumn{11}{c}{(``$\surd$'' if the protocol satisfies the property, ``$\times$'' if not)}
\end{tabular}
\end{table*}

IRONMAN \cite{IRONMAN-77} is one of the first social trust-based routing mechanism in which the nodes initially assign the highest trust value to their social friends. Then, encountered nodes exchange the history of their sent and received messages with each other, based on which they decrease the trust level of each other for each detected dropping message. Besides, a node increases the trust level of its encountered node when it receives a relaying message from that node. Additionally, the encountered nodes exchange their opinion about the trust level of other nodes with each other.
However, IRONMAN initially assigns the highest trust score to each node, and thus selfish nodes have a chance to only forward their messages selfishly until their reputation is higher than a threshold value.
To deal with this problem, in SENSE \cite{SENSE-66}, nodes' social features, battery level, and message hop count are used to identify their altruism. Next, two encountered nodes agree to calculate the reputation of each other if their battery level is above a threshold value. Then, if the nodes deduce that they are non-selfish to each other, they exchange the history of their sent and received messages as well as their opinion about the reputation of other nodes with each other to faster detect selfish nodes.

The authors in \cite{DynTrustRouting2014} propose a dynamic social trust management mechanism to secure and optimize DTN routing in which the combination of quality-of-service (QoS) trust and social trust are used to select trustworthy relay nodes. While the delivery probability is considered to measure the QoS trust, healthiness and unselfishness metrics are introduced to measure the nodes' social trust level. When two nodes contact each other, they calculate the trust value of each other based on their direct contact and indirect trust information. The experiments using a stochastic Petri Net technique demonstrate that this method outperforms some existing trust-based and non-trust-based DTN routing protocols in terms of data delivery and delay. Similarly, trust routing based on social similarity (TRSS) \cite{SecureSocSimil-112} incorporates the concept of social trust into DTN routing where the nodes' common interests and social similarities are used to quantify their trust level. Next, nodes with higher social trust levels are selected as the message relays. Chen \textit{et al.} \cite{SocialIoTTrust2016} use the concept of honesty, cooperativeness, and community-interest to establish social trust relations between nodes, based on which a social-aware application can adjust the best trust-related parameters not only for establishing secure communications but also maximizing the network performance.

While the studies above focus on DTNs, a number of recent studies have investigated the role of social trust in D2D communications. Ometov \textit{et al.} \cite{TrustedSocD2DOmetov} explore how the combination of human social-awareness and D2D communications can improve the communications performance and service quality. In particular, they propose a social-aware trusted D2D data delivery framework in which a coalitional game approach is employed to cluster mobile nodes based on their social tie strength and the degree of proximity. The evaluation results demonstrate that the proposed framework outperforms traditional cellular-only and network-assisted D2D communications in terms of energy efficiency and degree of connectivity. Similarly, Chen \textit{et al.} \cite{ExploitSocTieD2D2015} propose a coalitional game model to establish efficient and secure D2D cooperative communications by leveraging social trust and social reciprocity. The experiments show that this approach achieves up to 122\% performance gain in comparison with the cellular-only communications. Similar coalition formation solutions are proposed in \cite{TrustIoT2016}\cite{SocialTrsutCoalition2015} to establish social trust-based network-assisted D2D communications.

In addition to the coalition formation methods discussed above, some other solutions have been proposed to select trustworthy relay nodes in D2D communications. Zhang \textit{et al.}\cite{StoppingGameD2D} propose a stopping theory to identify effective and trustworthy relay nodes in D2D communications wherein the nodes' social and physical information is captured to establish social trust relations among them. The experiments demonstrate that the proposed scheme achieves up to 120\% and 45\% performance gain over the case without D2D cooperation and random relay selection, respectively. Yan \textit{et al.} \cite{PartnerSelection2017} propose a trust-oriented partner selection mechanism in D2D communications in which multi-dimensional trust relations between the sender and possible relay nodes is established by evaluating their cognition, emotion, and behavior trust. Next, a rough set decision-making algorithm is designed to choose the most reliable relay node.

\textbf{Summary:} Table \ref{tableWatchdog} summarizes the important characteristics of the watchdog and social trust-based systems in WRNs. It can be seen that almost all the watchdog systems rely on nodes' contact history, while the impact of the nodes' social relationships and preferences on the efficiency and effectiveness of watchdog systems are not explored sufficiently. In contrast to the watchdog mechanisms, the social trust-based systems exploit nodes' contact history and social relationships to choose more reliable and trustable relay nodes in message forwarding. Comparatively, most of the social trust-based systems in DTNs are fully distributed, whereas the social trust-based systems in D2D communications mainly take advantages of the underlying cellular network to establish the trust relationships between nodes and isolate selfish nodes. Besides, the majority of the social trust-based systems in DTNs analyze the impact of nodes' malicious behavior (\textit{e.g.,} disseminating false positives and false negatives) on the performance of selfish node detection. While, the social trust-based systems in D2D communications cannot protect the network against malicious nodes' misreporting or other trust-related attacks.

\subsection{Malicious Node Attack Detection Mechanisms}

\noindent In Section \ref{NodeAttackModels}, we introduced different types of attacks that can be launched by malicious nodes in WRNs. In this subsection, we discuss well-known attack detection mechanisms.

\subsubsection{Blackhole and Greyhole Detection Methods}
blackhole and greyhole are two common node attacks where an adversary drops all or a fraction of its relaying messages but forges its routing metrics to hide its malicious behavior. Although various blackhole and greyhole attack detection countermeasures have been proposed in wireless ad hoc networks (\textit{e.g.}, \cite{Ack-28} \cite{AttackMANETs}), they rely on end-to-end node connectivity that may not be applicable to WRNs.

Blackhole and greyhole detection mechanisms in WRNs primarily investigate the consistency of nodes' contact history and message exchange records to secure their communications and prevent the attackers from distributing falsified connectivity metrics. The authors in \cite{Thwarting-30} use encounter tickets to detect blackhole attackers in which two nodes sign an encounter ticket using their trusted private key identification when they contact each other. Accordingly, encountered nodes are required to submit their encounter tickets to their next encounters that prevent the attackers from claiming non-existing encounters. However, an adversary can still launch advanced types of the blackhole attack, such as tailgating wherein a node deliberately increases its contact frequency with popular nodes to attract more messages. To combat such attacks, a ticket-based prediction technique is designed in which a node predicts the competency of its encountered node to decide whether to forward a message to it or not.

While \cite{Thwarting-30} investigates the contact history of nodes to detect blackhole attackers in DTNs, Dini and Duca \cite{RepuBlack-75} propose a reputation system where selfish nodes disseminate reputation value 0 to never be chosen as a relay node, whereas misbehaving nodes disseminate reputation value 1 to attract more messages. When a node receives a message, it updates the reputation of all nodes that the message relayed throughout. To cope with misbehaving nodes, a survival model is used in which a node periodically decreases the reputation of other nodes if it does not receive a message from them within a time period. In addition, Li and Cao \cite{MitMisbehave-36} propose a detection system wherein a node is required to share the list of its sent and received messages with its next encountered node to help them judge whether this node has dropped any message or not. However, malicious nodes may manipulate their contact records to avoid being detected. To deal with this problem, a node is required to share a part of its contact records with other nodes, based on which the nodes can analyze the consistency of contact records received from different nodes and detect misreporting attackers.

While the methods discussed above can only detect blackhole attackers, Alajeely \textit{et al.} \cite{CatabolismAtt-82} introduce a new type of greyhole attack called Catabolism attack in which adversary nodes drop some received messages and inject new fake messages instead. To deal with this attack, a defense mechanism called Anabolism is proposed where a hash chain model is applied to detect the malicious nodes. Furthermore, Diep and Yeo \cite{ColluBlakGrey-34} propose a statistical defense scheme, namely SDBG, to detect both individual and colluding blackhole and greyhole attackers. To detect the individual attackers, encountered nodes are required to exchange their contact history that let the other nodes judge their behavior. In particular, some sort of forwarding ratio metrics are designed in SDBG that help a judging node to compare the routing behavior of a judged node against threshold values. If the judged node is detected as an individual attacker, SDBG starts detecting possible colluding attackers in two phases. In the first phase, judging node identifies the potential colluders with the judged node based on the number of their received messages from the judged node. In the second phase, the judging node uses the forwarding ratio metrics to investigate the number of messages the judged node has forward to the suspicions colluders. The simulation results illustrate that SDBG outperforms the method in \cite{MitMisbehave-36} with a detection rate of at least 70\%.

Saha \textit{et al.} \cite{lightDoS-69} discuss that exchanging table-based information between nodes can cause high communication cost and long detection time. Thus, they use special trusted nodes (TNs) with long-range connectivity over the SnW protocol to detect malicious nodes by addressing the questions what information should be exchanged between TNs and how often. To this aim, they consider three scenarios: (1) TNs only exchange their contact information; (2) TNs exchange the information of malicious nodes; and (3) TNs exchange the information of malicious nodes along with additional information. The experimental results demonstrate that scenario 2 reduces the detection time by 26\%, cost by 6\%, and the detection ratio by 15-25\% as compared to scenario 1. In addition, scenario 3 reduces the detection time by 45\% and the detection ratio by 10\% with a slight increase in cost as compared to scenario 2.

\subsubsection{Data Flooding Attack Detection Methods}
the primary goal of a flood attacker is to generate as many messages as possible to congest the network and waste the resources of other nodes. While several studies have attempted to alleviate the flood attack in wireless ad hoc networks \cite{FloodAttackAdHoc} and peer-to-peer networks \cite{SF-DRDoS-40}, they cannot be applied to WRNs because they require a permanent centralized monitoring server or end-to-end path information.

Recently, a number of studies have addressed the data flooding attack in WRNs. Li \textit{et al.} \cite{LieComply-43} study the impact of flood attack on the performance of single-copy and multi-copy DTN routing protocols and show that the data flooding attack can waste more than 80\% of the transmissions generated by honest nodes in the presence of 5\% of flooding attackers. To deal with the flood attackers, a rate-limiting method is proposed in which a node can replicate a limited number of message copies. However, counting all the number of messages generated by a particular node may not be possible in WRNs because of the lack of a centralized center. Hence, a claim-carry-and-check method is adopted where each node claims the number of its generated or replicated messages to other nodes. Thus, the other nodes can cross-check their carried claims to detect inconsistent claims. Diep and Yeo \cite{FloodAttack2017} propose an encounter-based mechanism to detect flooding attackers without imposing strict limitations on nodes' message generation rate. In particular, a burst-limit policy is applied to restrict the flooding attack where the nodes' normal message generation pattern is still controlled using a rate-limiting method, but they are still allowed to have a small and short burst of new messages. To this aim, encountered nodes are required to exchange the list of their send and received messages with each other that can help them judge if another node violates the burst-limit policy.

While the above-mentioned studies rely on nodes' contact history, Parris \textit{et al.} \cite{FriFlood-44} propose a social-based defense mechanism against flooding attackers wherein each node is required to sign its forwarding messages and attach the list of its friends in each message. Thus, the trusted social friends of the source node only can carry its messages. Nevertheless, an attacker may spoof the header of a message to falsely make its encountered node believes that it is relaying its friend's message. In a worse case, the attacker may spoof multiple MAC-layer addresses to replicate a huge number of messages to a particular node. To deal with these attacks, a key distribution mechanism is designed in which a message is discarded if it is not truly signed by a friend.

\subsubsection{Wormhole Attack Detection Methods}
a wormhole attacker receives messages at one location of the network and then tunnels and retransfers all or some of them to nodes at another location in the network. In this way, the wormhole attacker can disturb and manipulate the topology views of the network. While several recent studies have addressed the wormhole attack in traditional wireless ad hoc networks (\textit{e.g.}, \cite{WormholeAdHoc2008}), a limited number of works have addressed the wormhole attack in WRNs. Ren \textit{et al.} \cite{WormDTN-48} propose a geographical-based mechanism where the node mobility is utilized to detect a forbidden topology. In this method, mobile nodes reduce their transmission range for short time and then the nodes' geometric relations are analyzed to detect wormhole attacks. The evaluation results demonstrate that the detection ratio goes up as the network density increases. Furthermore, it is found that the detection ratio increases when nodes have higher mobility. The authors in \cite{StatWorm-49} propose a statistical-based approach in which infrastructure-based nodes collect and analyze the contact information of mobile normal nodes to detect and localize wormhole attackers. The detection process includes two phases: training and test. In the training phase, the average number of contacts between nodes over a period of time is calculated. Next, the testing phase checks if the ratio between the current node contacts and the mean contact number exceeds a threshold value.

\subsubsection{Sybil Attack Detection Methods}
a Sybil attacker (or Sybil) generates a large number of bogus identities or location information to establish many fake links in the network. Several detection techniques have been proposed for wireless networks that primarily use social network information (\textit{e.g.}, \cite{SybilLimit-58}) or cryptography techniques (\textit{e.g.}, \cite{LSR-59}) to detect Sybil attackers \cite{SybilIoT-55}. Nevertheless, detecting Sybils and establishing a global trust in WRNs entails major challenges due to various reasons, such as the poor knowledge of nodes about the network's global state.

\begin{table*}[!t]
%% increase table row spacing, adjust to taste
%\renewcommand{\arraystretch}{1.3}
% if using array.sty, it might be a good idea to tweak the value of
% \extrarowheight as needed to properly center the text within the cells
\caption{Summary of the attack detection mechanisms in wireless relay networks}
\label{tableAttackDetect}
\centering
\linespread{1.1}\selectfont
%% Some packages, such as MDW tools, offer better commands for making tables
%% than the plain LaTeX2e tabular which is used here.
\begin{tabular}{|*{12}{c|}} % repeats {c|} X times
\hline

\multirow{8}{*}{\rotatebox[origin=c]{90}{\textbf{Attack model}}} & 	
%\multirow{8}{*}{\rotatebox[origin=c]{90}{\textbf{Reference}}} &	
\multirow{8}{*}{\parbox[t]{13mm}{\textbf{Reference}}} &
\multirow{8}{*}{\parbox[t]{38mm}{\textbf{Principle of proposed solutions}}} &
\multicolumn{2}{c|}{\parbox[t]{8mm}{\textbf{Attack model}}} &
\multicolumn{4}{c|}{\parbox[t]{16mm}{\textbf{Detection \\information}}} &
\multirow{8}{*}{\rotatebox[origin=c]{90}{\textbf{Fully distributed}}} &	
\multirow{8}{*}{\parbox[t]{45mm}{\textbf{Specialties (+) and limitations (-)}}}  \\ [1em]
\cline{4-9}
&&&
\parbox[t]{1mm}{\multirow{6}*{\rotatebox[origin=c]{90}{\textbf{Individual}}}}	&
\parbox[t]{1mm}{\multirow{6}*{\rotatebox[origin=c]{90}{\textbf{Colluding}}}}  &

\parbox[t]{1mm}{\multirow{6}*{\rotatebox[origin=c]{90}{\textbf{Contact records}}}}  & \parbox[t]{1mm}{\multirow{6}*{\rotatebox[origin=c]{90}{\textbf{Message records}}}} &
\parbox[t]{1mm}{\multirow{6}*{\rotatebox[origin=c]{90}{\textbf{Social features}}}}  &
\parbox[t]{1mm}{\multirow{6}*{\rotatebox[origin=c]{90}{\textbf{Cryptography}}}} && \\

\multicolumn{2}{|c|}{}&&&&&&&&& \\ \multicolumn{2}{|c|}{}&&&&&&&&& \\
\multicolumn{2}{|c|}{}&&&&&&&&& \\ \multicolumn{2}{|c|}{}&&&&&&&&& \\  \multicolumn{2}{|c|}{}&&&&&&&&& \\
%-----------------------------------------
\midrule
\hline
\parbox[t]{2mm}{\multirow{15}{0.2in}{\rotatebox[origin=c]{90}{Blackhole and Greyhole Attacks}}} &
\multirow{3}{0.7in}{Li \textit{et al.} \cite{Thwarting-30}}&	
\multirow{3}{1.7in}{A contact ticket-based scheme to detect message droppers}&	
\multirow{3}*{$\surd$}	&
\multirow{3}*{$\times$}&	
\multirow{3}*{$\surd$}	&
\multirow{3}*{$\times$}&	
\multirow{3}*{$\times$}	&
\multirow{3}*{$\surd$}&
\multirow{3}*{$\surd$}	&
\multirow{3}{2in}{+ Predicts the node competency based on a belief system
\\- No consideration of colluding attacks}  \\ &&&&&&&&&& \\ &&&&&&&&&& \\  \cline{2-11}

&
\multirow{3}{0.7in}{Dini and Duca \cite{RepuBlack-75}}&	
\multirow{3}{1.7in}{The integration of reputation and probabilistic routing to detect attackers}&	
\multirow{3}*{$\surd$}	&
\multirow{3}*{$\times$}&	
\multirow{3}*{$\times$}	&
\multirow{3}*{$\surd$}&	
\multirow{3}*{$\times$}	&
\multirow{3}*{$\surd$}&
\multirow{3}*{$\surd$}	&
\multirow{3}{2in}{+ Designs an aging method to determine the nodes' reputation
\\- The lack of analytical evaluations} \\ &&&&&&&&&& \\ &&&&&&&&&&  \\ \cline{2-11}
&
\multirow{3}{0.7in}{Li and Cao \cite{MitMisbehave-36}}&	%check****
\multirow{3}{1.7in}{A method that checks the consistency of contact records to detect message droppers}&	
\multirow{3}*{$\surd$}	&
\multirow{3}*{$\surd$}&	
\multirow{3}*{$\surd$}	&
\multirow{3}*{$\surd$}&	
\multirow{3}*{$\times$}	&
\multirow{3}*{$\surd$}&
\multirow{3}*{$\surd$}	&
\multirow{3}{2in}{+ Can detect colluding misreporting nodes
\\- No evaluation of false positives and negatives} \\ &&&&&&&&&& \\ &&&&&&&&&&  \\ \cline{2-11}
&

\multirow{3}{0.7in}{Alajeely \textit{et al.} \cite{CatabolismAtt-82}}&	
\multirow{3}{1.7in}{A hash chain model to detect nodes that drop messages or inject fake messages}&	
\multirow{3}*{$\surd$}	&
\multirow{3}*{$\times$}&	
\multirow{3}*{$\times$}	&
\multirow{3}*{$\surd$}&	
\multirow{3}*{$\times$}	&
\multirow{3}*{$\times$}&
\multirow{3}*{$\surd$}	&
\multirow{3}{2in}{+ Introduces a new attack model
\\- Weak simulation settings} \\ &&&&&&&&&&  \\  &&&&&&&&&&  \\   \cline{2-11}
&
\multirow{2}{0.7in}{Pham and Yeo \cite{ColluBlakGrey-34}}&	
\multirow{2}{1.7in}{A statistical method to detect individual and colluding droppers}&	
\multirow{2}*{$\surd$}	&
\multirow{2}*{$\surd$}&	
\multirow{2}*{$\surd$}	&
\multirow{2}*{$\surd$}&	
\multirow{2}*{$\times$}	&
\multirow{2}*{$\surd$}&
\multirow{2}*{$\surd$}	&
\multirow{2}{2in}{+ Considers different contact manipulation models}  \\ &&&&&&&&&& \\  \cline{2-11}
&
\multirow{3}{0.7in}{Saha \textit{et al.} \cite{lightDoS-69}}&	
\multirow{3}{1.7in}{A lightweight detection scheme based on some trusted nodes}&	
\multirow{3}*{$\surd$}	&
\multirow{3}*{$\times$}&	
\multirow{3}*{$\surd$}	&
\multirow{3}*{$\times$}&	
\multirow{3}*{$\times$}	&
\multirow{3}*{$\surd$}&
\multirow{3}*{$\times$}	&
\multirow{3}{2in}{+ Achieves a better trade-off between the detection time and overhead
\\- relies on long-range wireless connections} \\ &&&&&&&&&& \\ &&&&&&&&&& \\
%-----------------------------------------------------------------------
\hline
\parbox[t]{2mm}{\multirow{9}{0.2in}{\rotatebox[origin=c]{90}{Flooding Attack}}} &

\multirow{3}{0.7in}{Li \textit{et al.} \cite{LieComply-43}}&	
\multirow{3}{1.7in}{A rate-limiting method to detect inconsistent node claims about the number of replicated messages}&	
\multirow{3}*{$\surd$}	&
\multirow{3}*{$\surd$}&	
\multirow{3}*{$\surd$}	&
\multirow{3}*{$\surd$}&	
\multirow{3}*{$\times$}	&
\multirow{3}*{$\surd$}&
\multirow{3}*{$\surd$}	&
\multirow{3}{2in}{+ Less communication, computation, and storage costs
\\- No comparison with previous work} \\ &&&&&&&&&& \\ &&&&&&&&&&  \\  \cline{2-11}
&
\multirow{3}{0.7in}{Diep and Yeo \cite{FloodAttack2017}}& %check****	
\multirow{3}{1.7in}{A rate-limiting method to detect flooding attacks that allows legitimate burst traffic}&	
\multirow{3}*{$\surd$}	&
\multirow{3}*{$\times$}&	
\multirow{3}*{$\surd$}	&
\multirow{3}*{$\surd$}&	
\multirow{3}*{$\times$}	&
\multirow{3}*{$\surd$}&
\multirow{3}*{$\surd$}	&
\multirow{3}{2in}{+ Can detect the burst traffic violation
\\- The lack of analytical evaluations} \\ &&&&&&&&&&  \\ &&&&&&&&&&  \\ \cline{2-11}

&
\multirow{3}{0.7in}{Parris and Henderson \cite{FriFlood-44}}&	
\multirow{3}{1.7in}{A social-based authentication system to detect flooding attacks}&	
\multirow{3}*{$\surd$}	&
\multirow{3}*{$\times$}&	
\multirow{3}*{$\surd$}	&
\multirow{3}*{$\surd$}&	
\multirow{3}*{$\surd$}	&
\multirow{3}*{$\surd$}&
\multirow{3}*{$\surd$}	&
\multirow{3}{2in}{+ Considers various attack models
\\- Evaluation with only one attacker} \\ &&&&&&&&&&  \\ &&&&&&&&&&  \\
%%-----------------------------------------------------------------------
\hline
\parbox[t]{2mm}{\multirow{6}{0.2in}{\rotatebox[origin=c]{90}{Wormhole}}} &
\multirow{2}{0.7in}{Ren \textit{et al.} \cite{WormDTN-48}}&	
\multirow{2}{1.7in}{A geographical method to exploit the presence of a forbidden topology}&	
\multirow{2}*{$\surd$}	&
\multirow{2}*{$\surd$}&	
\multirow{2}*{$\times$}	&
\multirow{2}*{$\times$}&	
\multirow{2}*{$\times$}	&
\multirow{2}*{$\times$}&
\multirow{2}*{$\surd$}	&
\multirow{2}{2in}{+ A fully distributed detection method
\\- Detection needs at least three nodes} \\ &&&&&&&&&& \\  \cline{2-11}
&
\multirow{3}{0.7in}{Pham and Yeo \cite{StatWorm-49}}&	
\multirow{3}{1.7in}{A statistical analysis method to detect and localize wormhole attackers}&	
\multirow{3}*{$\surd$}	&
\multirow{3}*{$\surd$}&	
\multirow{3}*{$\surd$}	&
\multirow{3}*{$\times$}&	
\multirow{3}*{$\times$}	&
\multirow{3}*{$\times$}&
\multirow{3}*{$\times$}	&
\multirow{3}{2in}{+ Detection mechanism does not rely on the number of nodes
\\- relies on infrastructure nodes} \\ &&&&&&&&&& \\ &&&&&&&&&&  \\
%%---------------------------------------------------------------------------------
\hline
\parbox[t]{2mm}{\multirow{16}{0.2in}{\rotatebox[origin=c]{90}{Sybil Attack}}} &
\multirow{3}{0.7in}{Trifunovic \textit{et al.} \cite{StalkLie-83}}&	
\multirow{3}{1.7in}{Study various types of Sybil attacks and evaluating their effectiveness}&	
\multirow{3}*{$\surd$}	&
\multirow{3}*{$\times$}&	
\multirow{3}*{$\surd$}	&
\multirow{3}*{$\times$}&	
\multirow{3}*{$\surd$}	&
\multirow{3}*{$\times$}&
\multirow{3}*{$\surd$}	&
\multirow{3}{2in}{+ The valuation of four benchmark Sybil defence systems
\\- No consideration of colluding attackers} \\ &&&&&&&&&& \\ &&&&&&&&&&  \\ \cline{2-11}
&
\multirow{2}{0.7in}{Liang \textit{et al.} \cite{TruEvalMSNs-54}}&	
\multirow{2}{1.7in}{A trustworthy Sybil-resisted system to detect the service review attacks}&	
\multirow{2}*{$\surd$}	&
\multirow{2}*{$\surd$}&	
\multirow{2}*{$\surd$}	&
\multirow{2}*{$\times$}&	
\multirow{2}*{$\surd$}	&
\multirow{2}*{$\surd$}&
\multirow{2}*{$\surd$}	&
\multirow{2}{2in}{+ Resists the review attacks without relying on a third authority} \\ &&&&&&&&&&   \\ \cline{2-11}
&
\multirow{3}{0.7in}{Sun \textit{et al.} \cite{DefSybil-53}}&	
\multirow{3}{1.7in}{A security mechanism against attackers that report forged virtual locations}&	
\multirow{3}*{$\surd$}	&
\multirow{3}*{$\times$}&	
\multirow{3}*{$\times$}	&
\multirow{3}*{$\times$}&	
\multirow{3}*{$\times$}	&
\multirow{3}*{$\surd$}&
\multirow{3}*{$\times$}	&
\multirow{3}{2in}{+ measures metrics in client side but removes Sybils on the server side
\\- No evaluation results} \\ &&&&&&&&&& \\ &&&&&&&&&&  \\ \cline{2-11}

&
\multirow{3}{0.7in}{Quercia and Hailes \cite{FriFoes-50}}&	
\multirow{3}{1.7in}{A social-based Sybil detection mechanism based on node ranking}&	
\multirow{3}*{$\surd$}	&
\multirow{3}*{$\times$}&	
\multirow{3}*{$\times$}	&
\multirow{3}*{$\times$}&	
\multirow{3}*{$\surd$}	&
\multirow{3}*{$\surd$}&
\multirow{3}*{$\surd$}	&
\multirow{3}{2in}{+ Applying different ranking techniques
\\- The possible wrong detection of an honest node as a Sybil} \\ &&&&&&&&&& \\ &&&&&&&&&& \\ \cline{2-11}
&

\multirow{3}{0.7in}{Chang \textit{et al.} \cite{SybilDef-52}}&	
\multirow{3}{1.7in}{A gateway-breaking algorithm to remove suspicious attack edges with high centrality}&	
\multirow{3}*{$\surd$}	&
\multirow{3}*{$\times$}&	
\multirow{3}*{$\times$}	&
\multirow{3}*{$\times$}&	
\multirow{3}*{$\surd$}	&
\multirow{3}*{$\times$}&
\multirow{3}*{$\times$}	&
\multirow{3}{2in}{+ Each node carries small social profiles
\\- relies on a centralized server} \\ &&&&&&&&&&  \\ &&&&&&&&&&  \\ \cline{2-11}
&
\multirow{3}{0.7in}{Zhang \textit{et al.} \cite{ExpSybil-57}}&	
\multirow{3}{1.7in}{A social-based detection method based on nodes' abnormal contacts and pseudonym unstable behaviors}&	
\multirow{3}*{$\surd$}	&
\multirow{3}*{$\surd$}&	
\multirow{3}*{$\surd$}	&
\multirow{3}*{$\times$}&	
\multirow{3}*{$\surd$}	&
\multirow{3}*{$\surd$}&
\multirow{3}*{$\times$}	&
\multirow{3}{2in}{+ Detection of colluding Sybils
\\- relies on a server to store nodes' contact information} \\ &&&&&&&&&& \\ &&&&&&&&&&  \\

\hline
\parbox[t]{2mm}{\multirow{8}{0.2in}{\rotatebox[origin=c]{90}{Trust-related Attack}}} &

\multirow{3}{0.7in}{Chen \textit{et al.} \cite{DynTrustRouting2014}}&	%check***
\multirow{3}{1.7in}{A dynamic trust management mechanism that is resilient against major trust attacks}&	
\multirow{3}*{$\surd$}	&
\multirow{3}*{$\surd$}&	
\multirow{3}*{$\surd$}	&
\multirow{3}*{$\times$}&	
\multirow{3}*{$\surd$}	&
\multirow{3}*{$\times$}&
\multirow{3}*{$\surd$}	&
\multirow{3}{2in}{+ An application-level trust optimization technique to discard less trustworthy recommendations
%\\- No comparison with recent works
} \\ &&&&&&&&&& \\ &&&&&&&&&&  \\  \cline{2-11}
&
\multirow{3}{0.7in}{Chen \textit{et al.} \cite{SocialIoTTrust2016}}& %check****	
\multirow{3}{1.7in}{An adaptive social trust mechanism that deals with several trust-related attacks}&	
\multirow{3}*{$\surd$}	&
\multirow{3}*{$\surd$}&	
\multirow{3}*{$\surd$}	&
\multirow{3}*{$\times$}&	
\multirow{3}*{$\surd$}	&
\multirow{3}*{$\times$}&
\multirow{3}*{$\surd$}	&
\multirow{3}{2in}{+ Resilience against attacks even in extremely hostile environments
\\- No comparison with relevant methods} \\ &&&&&&&&&&  \\  &&&&&&&&&&  \\ \cline{2-11}

&
\multirow{3}{0.7in}{Yao \textit{et al.} \cite{SecureSocSimil-112}}&  %check	
\multirow{3}{1.7in}{A secure routing protocol that tolerates different trust-related attacks}&	
\multirow{3}*{$\surd$}	&
\multirow{3}*{$\surd$}&	
\multirow{3}*{$\surd$}	&
\multirow{3}*{$\times$}&	
\multirow{3}*{$\surd$}	&
\multirow{3}*{$\times$}&
\multirow{3}*{$\surd$}	&
\multirow{3}{2in}{+ Provides incentives for malicious nodes
\\- no detailed descriptions about protecting against trust-related attacks} \\ &&&&&&&&&&  \\ &&&&&&&&&&  \\

\hline

\multicolumn{11}{c}{(``$\surd$'' if the protocol satisfies the property, ``$\times$'' if not)}
%\multicolumn{11}{c}{(``$\surd$'' if the protocol satisfies the property, ``$\times$'' if not)}
%\multicolumn{12}{c}{(``$\surd$'' if the protocol satisfies the property, ``$\times$'' if not)}
\end{tabular}
\end{table*}

In general, Sybil detection methods in WRNs explore nodes' mobility and spatiotemporal information to detect Sybil attacks. The authors in \cite{StalkLie-83} consider a case where a Sybil intentionally encounters its targeted nodes to enhance its contact frequency and strengthen the weight of its social relationships with them. The experiments demonstrate that implementing successful Sybil attacks using mobility is costly for a mobile Sybil because it needs to invest several hours to infiltrate a community successfully. Liang \textit{et al.} \cite{TruEvalMSNs-54} propose a sybil-resisted trustworthy service evaluation system, called SrTSE, in which two types of Sybils can exist. First, users who put a bad review about a service provided by a vendor while it is good. Second, a vendor along with a group of users who put good reviews about a bad service to increase its reputation. To prevent these attacks, SrTSE assumes that a user can only put one review about a vendor in a short period of time. Thus, if a user puts several reviews with different pseudonyms about a vendor at a particular time, it will be considered as a Sybil. Sun \textit{et al.} \cite{DefSybil-53} propose a geographical mechanism to detect Sybils that report forged virtual locations. In particular, a two-dimensional coordination system is designed on the server side that obtains the set of Euclidean distances between nodes and generates a set of candidate nodes for routing. In case a Sybil node forges an unreal location, a high dimensional location is generated as output inferring that the node is a Sybil node.

A couple of existing works exploit nodes' social network information to detect Sybil attacks. MobID \cite{FriFoes-50} deals with Sybil attackers who may produce several fake identities but have a few real-life relationships. In MobID, a node enlists the identity of its encountered nodes in two small networks: the network of friends and network of foes. Thus, the node explores the social similarity between the friends and foes networks to decide whether an unknown contacting node launches a Sybil attack or not. Chang \textit{et al.} \cite{SybilDef-52} consider a community-based MSN where both Sybils and well-behaved nodes exist in the network. In addition, a local ranking system is employed to identify trust and distrust relations among nodes. Thus, a node stores two random social profiles: a trust profile and a distrust profile. When two strangers contact each other, they exchange their trust profile with each other to calculate the trust and distrust levels of each other, based on which they can decide whether another one is Sybil or not. Zhang \textit{et al.} \cite{ExpSybil-57} introduce a social-based Sybil detection scheme in which contact patterns and pseudonym behavior of nodes are investigated to detect Sybils. Since the storage and computational capabilities of mobile devices are limited, cloud servers are utilized to process the nodes' contact traces and detect Sybils.

\subsubsection{Social Trust-related Attack Detection Methods}
\label{SocialTrustAttack}
the social trust relationships between mobile nodes can be exploited to establish reliable and secure communications in WRNs. Nevertheless, malicious nodes falsify the trust level of their owns or their friends in order to attract more services or messages) but later refuse to provide the promised services. Besides, they can launch a colluding attack to spoil the good reputation of well-behaved nodes. In general, three major trust-related attacks in the context of WRNs have been considered in the literature: self-promoting attack where an adversary aims to promote its trust level, bad-mouthing attack in which an adversary ruins the trust level or reputation of other (well-behaved) nodes, and ballot-stuffing attacks wherein an adversary exaggerates the trust level of other malicious nodes. To deal with these attacks, the authors in \cite{DynTrustRouting2014} investigate the consistency of the encounter tickets received from different nodes based on a metric called healthiness social trust (that is the belief of a node whether another node is malicious or not) to identify the self-promoting attacks. Moreover, the consistency of trust recommendations provided by other nodes is checked to detect the bad-mouthing and ballot-stuffing attacks. Similarly, the authors in \cite{SocialIoTTrust2016}\cite{SecureSocSimil-112} check the consistency of nodes' direct and indirect trust recommendations to detect the trust-related attacks.

\textbf{Summary:} Table \ref{tableAttackDetect} summarizes the main features of the attack detection techniques. It can be seen that a few numbers of the blackhole and greyhole attack detection mechanisms (\textit{i.e.}, \cite{MitMisbehave-36} and \cite{ColluBlakGrey-34}) can protect the network against colluding message droppers. In addition, nodes' social features and relationships are not considered in the existing blackhole and greyhole attack detection mechanisms. Besides, a limited number of data flooding and wormhole attack detection mechanisms are proposed in WRNs where only \cite{FriFlood-44} uses nodes' social features to detect malicious attackers. In contrast, almost all the Sybil detection methods primarily take nodes' social relationships into account to detect Sybil attackers. Nevertheless, a few numbers of them (\textit{e.g.}, \cite{TruEvalMSNs-54} and \cite{ExpSybil-57}) can detect the colluding Sybil attackers. Furthermore, the trust-based attack detection mechanisms mainly apply nodes' contact and social information to identify both untrustworthy individual and colluding nodes. While almost all the attack detection schemes are designed for DTNs, detecting malicious nodes in D2D communications with respect to their specific characteristics needs further explorations.

\noindent
\section{Incentive Mechanisms}
\label{Incentive}

\noindent
Mobile nodes may not be willing to share their resources with each other and participate in data relaying unless an appropriate incentive is provided. However, designing an effective and fair incentive mechanism in WRNs is extremely challenging because mobile nodes do not have complete information about the network's global state. Furthermore, nodes with different resource constraints and preferences may require different types of incentives to cooperate with each other in data delivery. The ultimate goal of an incentive scheme is to make the rewarding mechanism incentive-compatible implying that a node obtains the highest reward when it has honest behavior. Broadly, existing incentive mechanisms can be classified into three categories: tit-for-tat (TFT)-based, reputation-based, and credit-based schemes. In the rest of this section, we study well-known incentive schemes in each category and characterize their main features.

\noindent
\subsection{TFT-based Incentive Mechanisms}

\noindent The main idea in TFT-based mechanisms is to force nodes to exchange the same number of messages during an opportunistic contact. In other words, TFT-based mechanisms aim to ensure that mobile nodes provide better forwarding services for cooperative nodes but avoid selfish nodes. Shevade \textit{et al.} \cite{IncentDTNs-70} propose a TFT mechanism for DTNs in which the concepts of generosity and contrition are employed to respectively overcome bootstrapping (\textit{i.e.}, who starts the cooperation) and exploitation (\textit{i.e.}, when another node exploits) problems. In this work, encountered nodes exchange their contact information periodically, based on which a source node can calculate the forwarding path of its messages. Finally, when a message is delivered to its destination, intermediate nodes in the delivery path are awarded. The simulation results show that the data delivery ratio increases up to 60\% in comparison with a fully cooperative scenario. Similarly, the authors in \cite{BarterTrade-71} propose a barter-based approach where encountered nodes exchange the list of their messages with each other. Next, they identify candidate messages and their forwarding priorities. Finally, they exchange their messages one by one until all of them are processed or their connection is lost. However, the message exchange methods in \cite{IncentDTNs-70} and \cite{BarterTrade-71} can cause deadlocks in case nodes do not have the same number of messages. Meanwhile, the value of messages is not considered in their incentive mechanism.

To deal with above-mentioned problems, MobiTrade \cite{MobiTrade-72} allows nodes to exchange messages if they do not have the same number of messages. In MobiTrade, the value of a message is identified based on the number of nodes that are interested in the message and nodes' cooperation degree. In addition, a buffer allocation technique is applied that helps a node to split its buffer for each channel based on its knowledge of future demand. Similarly, Zhou \textit{et al.} \cite{ConSub-73} propose a TFT-based content dissemination scheme for publish-subscribe systems in which the order of forwarding messages is identified based on a content utility function. Specifically, the utility of a message for a certain node is identified based on the number of nodes interested in the message, node contact probability, and the cooperation level of nodes.

\begin{table*}[!t]
%% increase table row spacing, adjust to taste
%\renewcommand{\arraystretch}{1.3}
% if using array.sty, it might be a good idea to tweak the value of
% \extrarowheight as needed to properly center the text within the cells
\caption{Summary of the Tit-For-Tat and reputation-based incentive mechanisms}
\label{TFTReputationIncentives}
\centering
\linespread{1.1}\selectfont
%% Some packages, such as MDW tools, offer better commands for making tables
%% than the plain LaTeX2e tabular which is used here.
\begin{tabular}{|*{12}{c|}} % repeats {c|} X times
\hline

\multirow{3}{*}{\rotatebox[origin=c]{90}{\textbf{Model}}} & 	
%\multirow{3}{*}{\rotatebox[origin=c]{90}{\textbf{Ref.}}} &	
\multirow{3}{*}{\parbox[t]{13mm}{\textbf{Reference}}} &
\multirow{3}{*}{\parbox[t]{40mm}{\textbf{Principle of proposed solutions}}} &
\multirow{3}{*}{\parbox[t]{38mm}{\textbf{Incentive objective}}} &
\multirow{3}{*}{\parbox[t]{45mm}{\textbf{Specialties (+) and limitations (-)}}}  \\

\multicolumn{2}{|c|}{}&&& \\ \multicolumn{2}{|c|}{}&&& \\

\midrule
%-----------------------------------------
\hline
\parbox[t]{2mm}{\multirow{22}{0.2in}{\rotatebox[origin=c]{90}{Tit-For-Tat}}} &
\multirow{3}{0.7in}{Shevade \textit{et al.} \cite{IncentDTNs-70}}&	
\multirow{3}{1.7in}{An incentive mechanism that incorporates the generosity and contrition into the routing}&	
\multirow{3}{1.6in}{Maximizing the individual utility of nodes }&	
\multirow{3}{2in}{+ Extensive evaluations using both synthetic and real-world traces
\\- No consideration of fairness}  \\ &&&& \\ &&&& \\  \cline{2-5}
&
\multirow{3}{0.7in}{Buttyán \textit{et al.} \cite{BarterTrade-71}}&
\multirow{3}{1.7in}{A barter game model to promote node cooperation}&	
\multirow{3}{1.6in}{A node can obtain a message if it gives a message in return}&	
\multirow{3}{2in}{+ Considers the value of messages
\\- Limited analytical results}  \\ &&&& \\ &&&& \\ \cline{2-5}
&
\multirow{3}{0.7in}{Krifa \textit{et al.} \cite{MobiTrade-72}}&
\multirow{3}{1.7in}{A mechanism that allows nodes to trade its relaying messages and buy its interested messages}&	
\multirow{3}{1.6in}{Maximizing the expected utility of each stored message for future encounters}&	
\multirow{3}{2in}{+  Provides a customized resource allocation strategy for each node}  \\ &&&&  \\ &&&& \\  \cline{2-5}
&
\multirow{4}{0.7in}{Zhou \textit{et al.} \cite{ConSub-73}}&
\multirow{4}{1.7in}{A content-based incentive mechanism that stimulates nodes to transmit their messages to interested nodes in publish/subscribe systems}&	
\multirow{4}{1.6in}{Maximizing the future trading value of a stored message}&	
\multirow{4}{2in}{+ Selects forwarding messages based on their value and nodes' cooperation level
\\- Applies a complex content matching in the message selection process}  \\ &&&&  \\ &&&& \\ &&&& \\ \cline{2-5}
&
\multirow{4}{0.7in}{Hsu and Duan \cite{SocialGroupTFTD2D}}&
\multirow{4}{1.7in}{An equal-reciprocal incentive mechanism for social group-based data sharing in D2D communications}&	
\multirow{4}{1.6in}{Maximizing the utility of nodes, which is the amount of contents a node uploads minus those it downloads from the network}&	
\multirow{4}{2in}{+ No need to compute nodes' sharing probabilities in advance
\\- Caching capacity of nodes is not considered}  \\ &&&&  \\ &&&& \\ &&&& \\ \cline{2-5}
&
\multirow{3}{0.7in}{Pu \textit{et al.} \cite{D2DFogging2016}}&
\multirow{3}{1.7in}{A cooperative task offloading and execution framework in D2D communications}&	
\multirow{3}{1.6in}{A node can use the resources of other nodes if it shares more resources with the others}&	
\multirow{3}{2in}{+ The framework is lightweight and operates dynamically according to the system's current information
}  \\ &&&&  \\ &&&& \\  \cline{2-5}
&
\multirow{4}{0.7in}{Mastronarde \textit{et al.} \cite{RelayNotRelay}}&
\multirow{4}{1.7in}{A supervised learning algorithm that help nodes adapt their cooperation policy in D2D message relaying}&	
\multirow{4}{1.6in}{Maximizing the utility of nodes, which is the difference between a node's message forwarding utility and energy consumption}&	
\multirow{4}{2in}{+ Considers different mobility and relay budget classes
}  \\ &&&&  \\ &&&& \\ &&&& \\
%-----------------------------------------------------------
\hline
\parbox[t]{2mm}{\multirow{10}{0.2in}{\rotatebox[origin=c]{90}{Reputation-based}}} &
\multirow{3}{0.7in}{Wei \textit{et al.} \cite{MobiGame-76}}&	
\multirow{3}{1.7in}{A game-theoretic scheme to stimulate nodes and resist attacks}&	
\multirow{3}{1.6in}{Maximizing the individual utility of nodes in message forwarding}&	
\multirow{3}{2in}{+ Considers both security and fairness
\\- Limited analytical and simulation results}  \\ &&&& \\ &&&& \\  \cline{2-5}
&
\multirow{4}{0.7in}{Bigwood and Henderson \cite{IRONMAN-77}}&
\multirow{4}{1.7in}{A social-based trust mechanism to identify node reputation}&	
\multirow{4}{1.6in}{Maximizing the reputation of nodes based on their cooperation history}&	
\multirow{4}{2in}{+ The establishment of trust relationships between the nodes using their social information
\\- The lack of theoretical analysis}  \\ &&&& \\ &&&& \\ &&&& \\ \cline{2-5}
&
\multirow{4}{0.7in}{Silva \textit{et al.} \cite{MobiCoop-78}}&
\multirow{4}{1.7in}{A generalized system to stimulate cooperation in mobile applications}&	
\multirow{4}{1.6in}{Maximizing the battery life of mobile devices}&	
\multirow{4}{2in}{+ Using real application prototypes in the evaluations
\\- Relies on a web service to manage the node reputation}  \\ &&&& \\ &&&& \\ &&&& \\

\hline
%\multicolumn{12}{c}{(``$\surd$'' if the protocol satisfies the property, ``$\times$'' if not)}
\end{tabular}
\end{table*}

A number of recent studies employ the TFT approach to promote the cooperation of nodes in D2D communications. Hsu and Duan \cite{SocialGroupTFTD2D} propose an equal-reciprocal mechanism for data sharing in D2D communications where D2D nodes are grouped based on their physical information. Next, each node in a group can share the same number of content with each other. The experiments show that this method not only guarantees the fairness in content sharing but also maximizes the individual utility of the nodes. Additionally, D2D Fogging \cite{D2DFogging2016} is a collaborative task offloading and execution mechanism in which a set of TFT resource constraints and an energy budget constraint is introduced to stimulate over-exploited and free-rider nodes to participate in data sharing. The TFT resource constraints ensure that a node can utilize the resources of other nodes if it shares more resources with the others. Furthermore, Lyapunov optimization methods are developed to minimize the energy consumption of D2D nodes with respect to those incentive constraints. The simulations demonstrate that the energy consumption of nodes reduces by 30-40\% in comparison to a case each node executes its tasks locally. Mastronarde \textit{et al.} \cite{RelayNotRelay} employ an online supervised learning algorithm that helps a node learn its cooperation policy and make a decision whether to relay messages received from other nodes or not. The experimental results reveal that the network achieves the highest performance when there exist many nodes with high energy resources to relay messages.
\noindent
\subsection{Trust and Reputation-based Incentive Mechanisms}

\noindent In the trust and reputation-based incentives, mobile nodes assign appropriate reputation to each other based on their direct trust relationships or indirect trust recommendations provided by other nodes. Eventually, better services are provided for nodes with high reputation or strong trust relationships. Under these circumstances, the nodes are stimulated to relay messages received from other nodes to gain enough reputation so that they can get help from other nodes. However, identifying the actual reputation of nodes in WRNs is challenging because the nodes cannot observe the behavior of each other thoroughly. Meanwhile, malicious nodes can manipulate their reputation for pretending that they have participated in data delivery.

MobiGame \cite{MobiGame-76} is a user-centric reputation system wherein a node submits the receipts of its relaying messages to the source and destination nodes to obtain credits. A message for an intermediate node can be a good bundle if the node can forward the message before it expires or a bad bundle if the message is close to being expired. It is assumed that both selfish and malicious nodes exist in the network where the selfish nodes do not return the relay evidence to the previous relay nodes. Meanwhile, the malicious nodes distribute bad bundles to other nodes to waste their resources. To establish a fair interaction, a game-theoretic model is designed in which the costs and utilities of forwarding and receiving good and bad bundles are analyzed using perfect Bayesian equilibrium. Similarly, IRONMAN \cite{IRONMAN-77} applies nodes' self-reported social network information to initialize their reputation. When two nodes $A$ and $B$ contact each other, they exchange their contact history, message-forwarding history, and the reputation of other nodes with each other that can help them update their opinions about the reputation of each other and other nodes. Once the reputation of one of them, say $A$, is less than a threshold value, $B$ discards messages received from $A$ until $A$ improves its reputation by relaying messages received from other nodes.

While the reputation mechanisms proposed in \cite{MobiGame-76} and \cite{IRONMAN-77} are fully distributed, MobiCoop \cite{MobiCoop-78} designs reputation-based incentives for hybrid DTNs in which nodes can contact each other through both service-oriented and opportunistic communications. Each node uses both direct and indirect observations to update the reputation of other nodes. In particular, three parameters including battery level, the Internet connectivity, and cooperation degree are used to calculate the reputation of a node. For example, the highest reputation value is awarded to a node that has a low battery level and access to the Internet, but it is still willing to cooperate with other nodes. However, MobiCoop depends on a centralized web service that may not be available in distributed DTNs.

\textbf{Summary:} Table \ref{TFTReputationIncentives} summarizes the main features of our studied TFT-based and reputation-based incentive mechanisms. It can be seen that the TFT mechanisms stimulate mobile nodes in both DTNs and D2D communications, whereas almost all the reputation-based schemes focus on promoting node cooperation in DTNs. Comparatively, TFT-based methods can work well when the network traffic is high, but they cannot provide fairness if encountered nodes do not have the same number of messages to exchange. Meanwhile, the message selection process in TFT-based mechanisms can affect their effectiveness in terms of the message delivery ratio significantly. In contrast, the performance of reputation-based mechanisms highly depends on the direct observations of nodes and the distribution of recommended-based reputations \cite{RepSys-143}.

\noindent
\subsection{Credit-based Incentive Mechanisms}

\noindent
Credit-based incentive schemes employ different forms of virtual credit to reward the cooperative nodes where the rewarding is commonly managed by a third-party credit clearance center (CCC). The idea is that a node is rewarded credit for relaying messages received from other nodes or sharing its resources with them where it can later use its credit to pay other nodes for achieving its own utilities. In this way, selfish nodes are not rewarded if they do not relay messages for others, and thus they cannot afford to buy the forwarding service of other nodes.

Broadly, existing credit-based incentive mechanisms can be categorized into two classes: game-theoretic and security-based mechanisms. The game-theoretic schemes aim to establish a win-win credit assignment situation among interacting nodes, whereas the security-based methods attempt to ensure the security of credit. In addition, there are some miscellaneous credit-based mechanisms that do not fall in the game-theoretic and security-based mechanisms. In the following, we present well-known credit-based mechanisms in each category.

\subsubsection{Game-theoretic Credit Mechanisms}
game-theoretic methods are widely applied to characterize the cooperations and competitions among rational mobile nodes with conflicting interests in wireless communications \cite{GameCoop-84} \cite{GameResource2014}. For example, a BS can set constraints on the transmission parameters in D2D communications so that mobile nodes compete or cooperate with each other to reuse the radio resources efficiently. In general, a game in WRNs consists of a set of players (\textit{i.e.}, mobile nodes and BSs), rules, strategies, and payoff (or utility) where each player chooses a strategy with the aim of maximizing its utility. The payoff is normally calculated based on the difference between the reward and cost of relaying a message (\textit{e.g.}, resource consumptions). Assuming that mobile nodes are selfish and rational, a binding agreement or equilibrium point should lead to a win-win situation where no player can improve its utility by unilaterally deviating from the equilibrium. In the rest of this subsection, we first study non-cooperative game-based credit schemes, followed by introducing the cooperative game-based credit schemes.

\textbf{Stackelberg Game-based Credit Schemes:} Stackelberg game is commonly played between a BS (\textit{i.e.}, leader) and mobile nodes (\textit{i.e.}, followers) in which the BS has an incentive to share a channel with some nodes if it is profitable. The BS decides the price, and the nodes choose the transmission power and channel given the charging price. The utility of the BS can be defined as its throughput plus the price it charges, whereas the utility of the nodes is the difference between its throughput and the cost that it pays to the BS for using the channel \cite{JointAlocation}.

Sugiyama \textit{et al.} \cite{IncenDTNSer-90} propose a two-stage Stackelberg game-based pricing scheme wherein a network operator decides how much it should pay to mobile nodes if they participate in data delivery. First, the operator announces the total reward and the minimum number of required participating nodes. Next, candidate nodes play the game to decide whether they want to cooperate in data forwarding or not. Finally, the operator shares the reward among the collaborative nodes if their number is higher than the required value. The cost of relaying a message by a node is identified based on its storage and energy consumptions, and their revenue is identified using a prospect theory. Finally, backward induction method is used to analyze the tradeoff between the cost and revenue and find the Nash equilibrium. Chen \textit{et al.} \cite{Cache2016Chen} model the interactions between a BS and mobile nodes as a Stackelberg game in which the BS aims to minimize its rewarding cost, and mobile nodes aim to maximize their utility by choosing an appropriate caching decision. Particularly, an iterative gradient algorithm is applied to find the Stackelberg equilibrium and maximize the utility of both the BS and nodes. A similar Stackelberg game-based incentive scheme is proposed in \cite{JointSpectrumPower2016}.

\begin{figure}[!t]
\centering
\includegraphics[width=3.4 in]{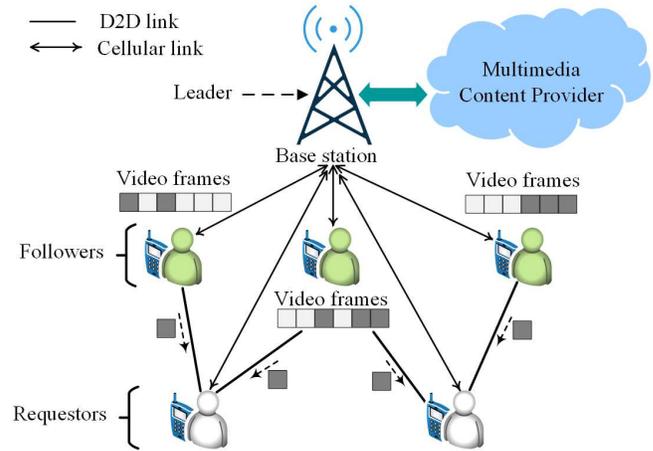}
\caption{An illustration of video content sharing via network-assisted D2D communication using Stackelberg game.}
\label{StackelbergVideoD2D}
\end{figure}

Some existing credit-based mechanisms employ Stackelberg game to optimize quality-driven multimedia video sharing in WRNs. Wu and Ma \cite{VideoIncen-93} propose an incentive scheme for distributing video files in which an interested node (leader) publishes a request to receive a video file by declaring a total credit for the delivery. Next, all the participants (followers) compete with each other to deliver the video file and earn credit. Since each video frame could increase the quality of the reconstructed video file, the destination node measures the value of each video frame using a utility function and reward each participating node based on its total contributions. The authors in \cite{QualityJointStackelberg2015} and \cite{SocialVideoStackelberg2017} employ Stackelberg game to design incentives for video sharing in network-assisted D2D communications in which the game is played between a multimedia content provider or BS as the leader and nodes with video contents as the followers (see Fig. \ref{StackelbergVideoD2D}). The objective is to maximize the benefits of the leader while stimulating the followers to participate in data sharing. Wang \textit{et al.} \cite{QualityJointStackelberg2015} propose a Stackelberg game-based source selection and power control solution where higher power and price are assigned to important packets to increase their delivery probability. Thus, the BS decides which devices to select and how much to pay for their provided radio resource. To this aim, Stackelberg equilibrium is employed to efficiently allocate the optimal power to the selected devices. Wu \textit{et al.} \cite{SocialVideoStackelberg2017} employ nodes' social and mobility features to select appropriate relay nodes that can efficiently distribute the video contents to interested peers. Next, a Stackelberg game is played between the BS and selected nodes to maximize their utility.

\textbf{Auction Game-based Credit Schemes:} auction is a popular incentive mechanism for scenarios in which the value of a service or trading item is undetermined. In a typical auction, a seller first announces the auction, and buyers respond to the auction in terms of biding. Next, the seller identifies the result of the auction and assigns the resources to the winners. Xu \textit{et al.} \cite{InterfAuction2012} propose a sequential second price auction to allocate spectrum resources in a network with a BS and multiple D2D devices where the spectrum resource units are auctioned off by D2D devices. In each round, the D2D devices offer a bid based on the value of the current resource unit, and then the BS allocates the unit to a device with the highest bit value but pays the second highest bid. The game continues until all the resource units are sold. The utility of a device is the difference between the total value of spectrum units obtained and the total payment. This work is extended in \cite{ResourceRevIte2013} where the game is played among a BS, cellular nodes, and D2D nodes. In particular, a reverse iterative combinatorial auction mechanism is modeled to efficiently allocate the spectrum resources and reduce the intra-cell interference wherein the buyers are motivated to offer multiple bids on combinations of resources iteratively and the seller asks the prices in each round. The experiments demonstrate that the system sum transmission rate increases as the number of D2D devices and resource units increases.

A major problem with the solutions in \cite{InterfAuction2012} and \cite{ResourceRevIte2013} is that D2D devices have to submit the game information (\textit{e.g.,} prices and costs) in each round of the game, while one of them will receive the reward finally that wastes their energy. To deal with this problem, Huang \textit{et al.} \cite{PostedPrice2016} propose a sequential posted pricing method in which the BS announces the auction by sending a posted price to the devices and assigns the resource unit to only one owner in each round. In this way, the BS stops activating the rest of devices because there exists already an active owner accepting the offer. The experiments show that this method achieves a better tradeoff between the BS's cost and the number of active devices.

Hajiesmaili \textit{et al.} \cite{HajiesmailiLaodD2D2017} propose an auction-based incentive scheme for load balancing in D2D-enabled cellular networks where the main goal is to dynamically shift the portion of the traffic of heavily-loaded cells to other under-utilized cells. To this aim, an online procurement auction mechanism is proposed in which multiple devices submit bids, and the BS evaluates the bids and purchases a subset of the resource units to satisfy the load balancing requirement while minimizing the social cost. The experiments demonstrate that the proposed scheme achieves a near offline-optimal performance.

\textbf{Bargaining Game-based Credit Schemes:} bargaining is a cooperative game approach in which the main goal is to fairly divide a certain surplus or credit among game players through negotiation. In the context of WRNs, bargaining game-based credit schemes have been extensively employed to model message trading between encountered mobile nodes with respect to their different criteria and preferences. Ning \textit{et al.} \cite{IncenDiss-85} consider a scenario in which mobile nodes willingly relay their interested messages but expect credit for relaying messages that they are interested.
%Thus, a node will be interested to relay messages with high delivery probability.
Since a credit is awarded only to the first deliverer and none of the nodes want to waste their resources, they design a two-player bargaining game where the encountered nodes negotiate over the value of their messages with respect to their delivery probability. Specifically, Nash bargaining equilibrium is employed to find an optimal solution, yielding the players exchange messages with the maximum gained credit. Similarly, self-interest-driven (SID)\cite{SID-92} proposes a two-player bargaining game for ad distribution wherein the players can trade both ad packets and virtual checks attached to each ad packet. In particular, the Nash bargaining equilibrium is applied to find a Pareto optimal point where both the players can reach a binding agreement. However, the rewarding mechanism in \cite{IncenDiss-85} and \cite{SID-92} are not fair because only the last-hop final deliverers are rewarded.

%considers an ad distribution scenario in which a source node attaches a virtual check to each ad packet when it replicates the ad to intermediate nodes. When an interested intermediate node receives the packet, it signs the check and authorizes the deliverer as the current owner of the check, based on which it can later request the ad provider to pay credit for each signed check.

Wu \textit{et al.} \cite{GameIncentProbRouting-87} propose a bargaining model to stimulate selfish nodes to cooperate in probabilistic routing protocols. The message trading is motivated by a marketing concept in which a message as a good is traded from a node with lower delivery probability to another node with a higher delivery probability. Thus, the current carrier of a message (seller) bargains with another node (buyer) over the value of the message in some rounds until an agreement on the price is reached, or they finally disagree. To identify the best strategy profile, a unique subgame perfect equilibrium is applied that helps the players to reach an agreement in the first round.

A number of bargaining schemes consider the sender of a message as a buyer who wants to buy the forwarding service of the receiver who is a seller. Li \textit{et al.} \cite{BargIncentive-86} design a two-player bargaining game assuming that the buffer and energy level of nodes are limited. First, the buyer offers price \textit{a} considering its free buffer, current wealth, and the message time-to-live (TTL). In contrast, the seller offers price \textit{b} with respect to its resources and the wealth. Next, they either agree to trade the message with price $\frac{(a+b)}{2}$ if $\mathrm{a}\mathrm{\ge }\mathrm{b}$ or disagree if $\mathrm{a<b}$. Furthermore, a bidding function is designed in a way that the buyer offers a high price when it is rich. In addition, the seller offers a high price when its resources are limited and a lower price when it is poor and needs to guarantee the forwarding of its own messages. Similarly, Jedari \textit{et al.} \cite{Jedari-98} propose an alternating-offers bargaining game, namely GISSO, in which the buyer and seller value the forwarding service based on their individual and social utilities where the utility of the messages is identified based on their social tie strength and message appraisal. Next, they negotiate over the service value in some rounds until they reach an agreement or the game is over. In GISSO, subgame perfect Nash equilibrium is applied to establish a win-win condition between the nodes where backward induction is employed to identify the best strategy for the players. Similar bargaining-based incentive schemes have been proposed in \cite{GameRelay-94} \cite{MultiPlayBarg-99}.

\begin{table*}[!t]
%% increase table row spacing, adjust to taste
%\renewcommand{\arraystretch}{1.3}
% if using array.sty, it might be a good idea to tweak the value of
% \extrarowheight as needed to properly center the text within the cells
\caption{Summary of the game-based credit mechanisms}
\label{IncentiveGame}
\centering

%\linespread{1.1}\selectfont
%% Some packages, such as MDW tools, offer better commands for making tables
%% than the plain LaTeX2e tabular which is used here.
\begin{tabular}{|*{6}{l|}} % repeats {c|} X times
\hline
\multicolumn{1}{|c|}{\parbox[t]{30mm}{\textbf{Game approach}}}
& \multicolumn{1}{|l|}{\parbox[t]{55mm}{\textbf{Incentive objective}}}
& \multicolumn{1}{|l|}{\parbox[t]{27mm}{\textbf{Analytical tools}}}
& \multicolumn{1}{|l|}{\parbox[t]{50mm}{\textbf{Achieved performance}}}
\\
\toprule

\hline

\multicolumn{1}{|l}{\parbox[t]{25mm}{Stackelberg game \\
\\Sugiyama \textit{et al.} \cite{IncenDTNSer-90} \\ Chen \textit{et al.} \cite{Cache2016Chen} \\ Yin \textit{et al.} \cite{JointSpectrumPower2016}}}
& \multicolumn{1}{|l|}{\parbox[t]{55mm}{
\cite{IncenDTNSer-90}-Attracting more nodes by the operator while minimize losing their energy and buffer
\\\cite{Cache2016Chen}-Minimizing the cost of BS while maximizing the utility of nodes
\\\cite{JointSpectrumPower2016}-Maximizing the sum rates of D2D devices while guaranteeing the cellular nodes' data rate requirement
\\ }}
& \multicolumn{1}{|l|}{\parbox[t]{27mm}{
\cite{IncenDTNSer-90}-Backward induction
\\\cite{Cache2016Chen}-Subgame perfect equilibrium
\\\cite{JointSpectrumPower2016}-Successive convex approximation}}

& \multicolumn{1}{|l|}{\parbox[t]{50mm}{
\cite{IncenDTNSer-90}-A win-win relationship between operators and mobile nodes
\\\cite{Cache2016Chen}-The caching scheme is beneficial to D2D devices if their requested pattern is more heterogeneous
\\\cite{JointSpectrumPower2016}-Achieves high performance while reducing the overhead of cellular nodes
\\
}} \\
%\midrule

\hline

\multicolumn{1}{|l}{\parbox[t]{25mm}{Auction game \\
\\Xu \textit{et al.} \cite{InterfAuction2012} \\ Xu \textit{et al.} \cite{ResourceRevIte2013} \\ Huang \textit{et al.} \cite{PostedPrice2016} \\ Hajiesmaili \textit{et al.} \cite{HajiesmailiLaodD2D2017}}}
& \multicolumn{1}{|l|}{\parbox[t]{55mm}{\cite{InterfAuction2012}-Maximizing the sum rate of the BS and the nodes' obtained resources while minimizing the nodes' payments
\\\cite{ResourceRevIte2013}-Maximizing the network sum rate by allowing cellular nodes to share their resources
\\\cite{ResourceRevIte2013}-Minimizing the overhead of the BS and the energy consumption of D2D devices
\\\cite{HajiesmailiLaodD2D2017}-Fulfilling load balancing requirement with the minimum social cost}}
& \multicolumn{1}{|l|}{\parbox[t]{27mm}{\cite{InterfAuction2012}-Subgame perfect equilibrium
\\\cite{ResourceRevIte2013}-Integer linear program
 \\\cite{PostedPrice2016}-Backward induction
 \\\cite{HajiesmailiLaodD2D2017}-Mixed integer linear program}}
& \multicolumn{1}{|l|}{\parbox[t]{50mm}{\cite{InterfAuction2012}-High performance on the system sum rate, efficiency, and fairness
\\\cite{ResourceRevIte2013}-Superior to the random allocation, high system efficiency, and stable over different parameters of nodes and resources
\\\cite{PostedPrice2016}-A better tradeoff between the BS's cost and the winning percentage of nodes
\\\cite{HajiesmailiLaodD2D2017}-Reduces the cost by 45\% compared with an alternative heuristic
\\
}} \\

\hline
\multicolumn{1}{|l}{\parbox[t]{25mm}{Bargaining game \\
\\Ning \textit{et al.} \cite{IncenDiss-85} \\ Ning \textit{et al.} \cite{SID-92} \\ Wei \textit{et al.} \cite{GameIncentProbRouting-87} \\ Li \textit{et al.} \cite{BargIncentive-86} \\ Jedari \textit{et al.} \cite{Jedari-98} \\ Xu \textit{et al.} \cite{GameRelay-94} \\ Li \textit{et al.} \cite{MultiPlayBarg-99}}}
& \multicolumn{1}{|l|}{\parbox[t]{55mm}{\cite{IncenDiss-85}-Maximizing the reward of sender nodes
\\\cite{SID-92}-Gaining a balanced credit while distributing as many ads as possible
\\\cite{GameIncentProbRouting-87}-Earning higher credit balance
\\\cite{BargIncentive-86}-Maximizing the node utility based on the buffer space and TTL
\\\cite{Jedari-98}-Maximizing the node utility considering the message TTL, delivery delay, and social tie
\\\cite{GameRelay-94}-Maximizing the node utility based on the buffer, energy, and TTL
\\\cite{MultiPlayBarg-99}-Saving the forward capability of nodes to serve the fitness messages}}
& \multicolumn{1}{|l|}{\parbox[t]{27mm}{\cite{IncenDiss-85}\cite{SID-92}\cite{MultiPlayBarg-99}-Nash bargaining theorem
\\\cite{GameIncentProbRouting-87}\cite{Jedari-98}\cite{GameRelay-94}-Subgame perfect equilibrium}}
& \multicolumn{1}{|l|}{\parbox[t]{50mm}{\cite{SID-92}-Reduces the transmission cost while maintaining a good delivery ratio and delay
\\\cite{GameIncentProbRouting-87}-Up to 75.8\% gain in data delivery in comparison with a non-incentive routing
\\\cite{BargIncentive-86}-Reduces the buffer consumption while delivering messages before the expiration
\\\cite{Jedari-98}\cite{GameRelay-94}-A good data delivery ratio and delay in the presence of selfish nodes
\\\cite{MultiPlayBarg-99}-Saves the network bandwidth and buffer while keeping a high delivery ratio
\\
}} \\

\hline
\multicolumn{1}{|l}{\parbox[t]{25mm}{Coalition game \\
\\ Han \textit{et al.} \cite{ZhuHanCure4Curse2009} \\ Akkarajitsakul \textit{et al.} \cite{EkramCOalition2013} \\ Zhang \textit{et al.} \cite{resourceD2D2013} \\ WZhu \textit{et al.} \cite{EnergyIncentiveCoalition2016} \\ Cao \textit{et al.} \cite{SocialTrsutCoalition2015} \\ Xiao \textit{et al.} \cite{OverlapCOalition2015} \\ Zhao and Song \cite{CoalitionGraph2017} \\ Wang \textit{et al.} \cite{CoalitionSocialCommunity2016}}}
& \multicolumn{1}{|l|}{\parbox[t]{55mm}{\cite{ZhuHanCure4Curse2009}-Establishes stable coalitions in which backbone and boundary nodes fairly cooperative
\\\cite{EkramCOalition2013}-Maximizing the nodes' payoffs
\\\cite{resourceD2D2013}-Nodes intend to maximize their utility, hence they have an incentive to form strong and stable coalitions
\\\cite{EnergyIncentiveCoalition2016}-Physically neighboring nodes form coalitions to minimize their energy consumption
\\\cite{SocialTrsutCoalition2015}-Maximizing the utility of spectrum sharing by stimulating nodes in a coalition to cooperate with each other
\\\cite{OverlapCOalition2015}-Each node chooses a specific BS to maximize its transmit rate per bandwidth price
\\\cite{CoalitionGraph2017}-Minimizing the power consumption of nodes while satisfying their power budget
\\\cite{CoalitionSocialCommunity2016}-BS maximizes the system sum rate while D2D devices maximize their individual payoffs}}
& \multicolumn{1}{|l|}{\parbox[t]{27mm}{\cite{ZhuHanCure4Curse2009}-Market fairness
\\\cite{EkramCOalition2013}-Markov chain model
\\\cite{resourceD2D2013}-Max-coalition order
\\\cite{OverlapCOalition2015}-Matching theory
\\\cite{CoalitionGraph2017}-Coalitional graph game
\\\cite{CoalitionSocialCommunity2016}-Defection function}}
& \multicolumn{1}{|l|}{\parbox[t]{50mm}{\cite{ZhuHanCure4Curse2009}-The network connectivity is improved by about 50\%
\\\cite{EkramCOalition2013}-Nodes achieve higher payoff comparing to a case they act alone
\\\cite{EnergyIncentiveCoalition2016}-All nodes participating in D2D content sharing achieve positive utilities
\\\cite{SocialTrsutCoalition2015}-Improves the nodes' perception quality of mobile video multicast effectively
\\\cite{OverlapCOalition2015}-Improves the system performance, especially in a large coverage area with a large number of D2D devices
\\\cite{CoalitionGraph2017}-Power consumption is almost optimal in a small-scale D2D network
\\\cite{CoalitionSocialCommunity2016}-Improves the system performance up to 93\% in compare to the case without community cooperation
\\
}} \\

\hline
\multicolumn{1}{|l}{\parbox[t]{25mm}{Algorithmic game \\ Cai \textit{et al.} \cite{IncentComp-97}
}}
& \multicolumn{1}{|l|}{\parbox[t]{55mm}{Maximizing the reward of only when they honestly report their encounter probability}}
& \multicolumn{1}{|l|}{\parbox[t]{27mm}{Sequential stopping rule}}
& \multicolumn{1}{|l|}{\parbox[t]{50mm}{Achieves higher data delivery ratio with low overhead
\\
}} \\

\hline
\multicolumn{1}{|l}{\parbox[t]{25mm}{Evolutionary game \\ \\ El-Azouzi \textit{et al.} \cite{EvoluEquili-89} \\ Lena Cota \textit{et al.} \cite{RACOON++} \\ Wang \textit{et al.} \cite{VPEFIncentive}
}}

& \multicolumn{1}{|l|}{\parbox[t]{55mm}{\cite{EvoluEquili-89}-Maximizing the probability of success
\cite{RACOON++}-Tolerating node selfishness while achieving high system performance
\\\cite{VPEFIncentive}-Maximizing sum of all nodes' utilities}}
& \multicolumn{1}{|l|}{\parbox[t]{27mm}{Evolutionary game theory}}
& \multicolumn{1}{|l|}{\parbox[t]{50mm}{\cite{EvoluEquili-89}-Reaches the equilibrium point using the nodes' local estimations
\\\cite{RACOON++}-Improves the bandwidth overhead by 22\% in a live streaming use case
}} \\

\hline
\multicolumn{1}{|l}{\parbox[t]{25mm}{Minority game \\ Chahin \textit{et al.} \cite{MinorityInc-91}
}}

& \multicolumn{1}{|c|}{\parbox[t]{55mm}{An optimal performance tradeoff between the delivery ratio and resource consumption}}
& \multicolumn{1}{|c|}{\parbox[t]{27mm}{Nash equilibrium}}
& \multicolumn{1}{|c|}{\parbox[t]{50mm}{Reaches the equilibrium point using the nodes' local estimations
}} \\ &&& \\

\hline
\multicolumn{1}{|l}{\parbox[t]{25mm}{Repeated game \\ \\ Huang \textit{et al.} \cite{RepeatHuang2016} \\ Barua \textit{et al.} \cite{RepeatBarua2016}
}}

& \multicolumn{1}{|l|}{\parbox[t]{55mm}{\cite{RepeatHuang2016}-Maximizing the payoffs of BSs as players, which are the payoffs from both cellular and D2D communications using radio resources
\\\cite{RepeatBarua2016}-Maximizing the utility of both BS and D2D devices in the presence of selfish nodes}}
& \multicolumn{1}{|l|}{\parbox[t]{27mm}{\cite{RepeatHuang2016}-Nash equilibrium derivations
\\\cite{RepeatBarua2016}-Nash equilibrium}}
& \multicolumn{1}{|l|}{\parbox[t]{50mm}{\cite{RepeatHuang2016}-Improves the system sum data rate and sum gain
\\\cite{RepeatBarua2016}-Maximizing the utility of the BS and D2D devices while resists selfish deviations
}} \\

\hline
\multicolumn{1}{|l}{\parbox[t]{25mm}{Mean filed game \\ Li \textit{et al.} \cite{MeanFieldLi2017}
}}

& \multicolumn{1}{|l|}{\parbox[t]{55mm}{Stimulate devices to truthfully reports the number of chunks they receive}}
& \multicolumn{1}{|l|}{\parbox[t]{27mm}{Mean field equilibrium}}
& \multicolumn{1}{|l|}{\parbox[t]{50mm}{Implementation on Android devices illustrates its efficient performance
}} \\

\hline
\multicolumn{1}{|l}{\parbox[t]{25mm}{Signaling game \\ Zhang \textit{et al.} \cite{SignalingD2DIncentive2017}
}}

& \multicolumn{1}{|l|}{\parbox[t]{55mm}{Maximizing the monetary benefit of nodes while guaranteeing a non-zero payoff for the BS}}
& \multicolumn{1}{|l|}{\parbox[t]{27mm}{Separating equilibrium}}
& \multicolumn{1}{|l|}{\parbox[t]{50mm}{Improves the system sum transmission rate
}} \\

\hline
\multicolumn{1}{|l}{\parbox[t]{25mm}{Network formation game \\ \\ Wang \textit{et al.} \cite{SocialD2DOffload}
}}

& \multicolumn{1}{|l|}{\parbox[t]{55mm}{Maximizing the individual payoffs of nodes}}
& \multicolumn{1}{|l|}{\parbox[t]{27mm}{Pairwise stability}}
& \multicolumn{1}{|l|}{\parbox[t]{50mm}{The performance gap between selfish and selfless nodes becomes smaller as the communication cost of cellular and D2D transmissions increases
\\
}} \\

\hline
%\multicolumn{12}{c}{(``$\surd$'' if the protocol satisfies the property, ``$\times$'' if not)}
\end{tabular}
\end{table*}

\textbf{Coalition Formation Game-based Credit Schemes:} coalition formation is a cooperative game approach in which a set of players (\textit{e.g.}, mobile nodes) agree to act as a single entity to gain a higher payoff, which is called coalition value. Han and Poor \cite{ZhuHanCure4Curse2009} study data forwarding in DTNs by highlighting that nodes on the boundary of the network (boundary nodes) are not willing to cooperate with backbone nodes in data relaying. To deal with this problem, the concept of \textit{core} is employed to establish stable coalitions in which the boundary and backbone nodes in a coalition have the incentive to cooperate with each other in data transmission. Next, they propose a routing protocol based on the coalition and repeated games, which improves the network connectivity by about 50\%. Similarly, Akkarajitsakul \textit{et al.} \cite{EkramCOalition2013} design a coalitional game to stimulate the cooperation of selfish nodes wherein the nodes decide either join or leave a coalition based on their individual payoffs. The individual payoff of a node is identified based on the delivery delay of their messages received from the BS and the cost incurred by this node for relaying the messages to other nodes. Using a Markov chain model to evaluate the stability of the coalitions, the experiments demonstrate that the nodes achieve a non-zero payoff.

A couple of coalitional game-based incentive schemes aim to design efficient content distribution and resource allocation protocols in D2D communications. Zhang \textit{et al.} \cite{resourceD2D2013} design a merge-and-split coalitional game with a transferable payoff (\textit{i.e.}, utilities like money are allocated to the players in the coalition) to efficiently allocate the spectrum resources between D2D and cellular devices. The utility of the D2D and cellular devices is defined as the sum transmission rate they can achieve through the resource blocks allocated to them. Hence, the game is divided into several sub-games where each sub-game addresses the resource allocation problem of one resource block. Since the nodes aim to maximize their utility, they have an incentive to form a strong group and win their preferred spectrum resources. In contrast to \cite{resourceD2D2013}, Zhu \textit{et al.} \cite{EnergyIncentiveCoalition2016} employ a non-transferable coalition formation game (\textit{i.e.}, different players have different interpretations of utilities) for energy-aware content sharing through D2D communication. Similarly, Xiao \textit{et al.} \cite{OverlapCOalition2015} model a Bayesian overlapping coalition game with non-transferable payoffs for efficient spectrum resource allocation.

Some studies exploit nodes' social features to form strong coalitions in D2D communications. The authors in \cite{SocialTrsutCoalition2015}\cite{CoalitionSocialCommunity2016} employ nodes' social tie information (\textit{e.g.}, social trust and reciprocity) to form stable coalitions, based on which D2D devices are stimulated to share their resources with each other, and the BS can share the spectrum resources efficiently. Similarly, Zhao \textit{et al.} \cite{CoalitionGraph2017} propose a coalition game-based incentive mechanism in D2D communications where the objective is to minimize the total power consumption while satisfying nodes' social incentive constraints.

%\vspace{0.5\baselineskip}
\textbf{Other Game-based Incentive Approaches:} other types of game-based credit mechanisms have been proposed in DTNs. Cai \textit{et al.} \cite{IncentComp-97} incorporate algorithmic game theory into the Two-hop protocol where a sequential stopping rule is employed to select the best relay nodes with maximum reward. Next, a second-price auction game is applied to identify the reward value in which a relay node can get the maximum reward if it reports its routing metrics honestly. Once a message is delivered to its destination, the source node rewards the intermediate nodes in the delivery path. The authors in \cite{EvoluEquili-89} employ an evolutionary game to promote the cooperation of nodes in the Two-hop protocol. Similarly, Wang \textit{et al.} propose a simple but effective incentive approach based on evolutionary game theory in community-based opportunistic networks wherein nodes voluntarily participate in message relaying and punish other non-cooperative nodes. In addition, an entry fee is received from nodes who want to participate in relaying messages in a community. The theoretical experiments prove that the efficiency loss of this scheme is $\frac{4}{8+M}$ where $M$ is the number of network nodes. Chahin \textit{et al.} \cite{MinorityInc-91} employ minority game to efficiently reward mobile nodes with respect to their mobility and resource consumption. The game aims to select a fraction of relay nodes (\textit{i.e.}, the minority) that are willing to participate in relaying a message on behalf of a source node under imperfect state information. The objective is to achieve an optimal performance tradeoff between the delivery probability and resource consumptions.

A number of other game-theoretic incentive approaches are proposed in D2D communications. Huang \textit{et al.} \cite{RepeatHuang2016} propose a repeated game for inter-cell scenarios where a D2D link is located in the overlapping area of two neighboring cells. In particular, the BSs are considered as the game players that compete for the resource demands of D2D devices where their utility is using the radio resources for both cellular and D2D communications. Barua \textit{et al.} \cite{RepeatBarua2016} design a repeated game for cooperative content sharing in which a D2D node receives contents from the BS and broadcasts them to interested nodes. Since selfish nodes do not cooperate in data forwarding, the game takes the nodes' cooperation level into account to select the best content carriers. While nodes with high cooperation level are rewarded by the BS, selfish nodes are punished in the next round of the game by giving their interested contents through cellular links. Li \textit{et al.} \cite{MeanFieldLi2017} design a mean filed game to encourage truth-telling about individual nodes states by paying monetary payments in a D2D real-time content streaming scenario. Furthermore, Zhang \textit{et al.} \cite{SignalingD2DIncentive2017} propose a signaling game-based incentive scheme for D2D content distribution wherein the main objective is to maximize the nodes' monetary profits while guaranteeing a non-negative utility for the BS.

\textbf{Summary:} Table \ref{IncentiveGame} summarizes the incentive objectives, analytical tools, and the major performance results of our studied game-theoretic credit schemes. It can be seen that the Stackelberg game-based incentive mechanisms mainly model the interactions between a BS, cellular nodes, and D2D nodes where the main objective is to efficiently allocate the spectrum resources, minimize the cost of the BS while maximizing the utility of cellular and D2D nodes. The auction game-based methods primarily aim to maximize the system sum transmission rate where the nodes can obtain maximum resource units with minimum payments. While the existing Stackelberg and auction game-based incentive approaches focus on D2D communications, the bargaining game-based mechanisms model message trading between mobile nodes where Nash bargaining and subgame perfect equilibrium solutions are mainly employed to find the equilibrium points. In addition, the coalition formation game solutions model content distribution in multi-hop cluster-based D2D communication where the main objective is to stimulate nodes inside the coalitions to participate with each other in data distribution. Furthermore, the other game-based incentive approaches aim to stimulate D2D devices to collaborate in data sharing with each other while allocating the spectrum resources efficiently.

\subsubsection{Security-based Credit Mechanisms}
Some existing studies incorporate security issues into credit mechanisms to protect them against various internal attacks (\textit{e.g.}, edge insertion and edge removal) in which malicious nodes strive to maximize their reward but reduce the reward of honest nodes.

SMART \cite{SMART-100} is a well-known secure pricing scheme in which the concept of \textit{layered coin} is employed to secure the rewarding and achieve fairness. First, the source of a message generates the first layer of the coin to indicate the credit value and rewarding policy. Next, each intermediate node adds a new layer to the coin by attaching its digital signature to show its participation in relaying the message. Once the message is delivered, nodes in the delivery path share the credit according to a profit-sharing model. However, malicious nodes may insert or remove fake layers or collude with each other to gain extra rewards. To overcome these attacks, a layer concatenation technique is designed in which the information of the previous and next layers are attached to the current layer to protect the layered coins against such attacks. Similarly, Lu \textit{et al.} \cite{Pi-102} employ the concept of layered coin to secure the credit assignment in which the source nodes reward the nodes in the delivery path of successfully delivered messages. To achieve fairness, nodes participated in relay a message obtain a reputation even if the message is not successfully delivered to its destination. Chen \textit{et al.} \cite{SecureCredit-107} introduce contribution time to reward relay nodes in the earliest delivery path of messages where the contribution time is the period of time between the receiving and forwarding of a message by a relay node. Using this method, a malicious node has no incentive to launch the edge insertion, removal, or content manipulate attacks because only nodes in the earliest delivery path receive credit.

Some security-based credit schemes aim at detecting layer insertion and removal attacks. Threshold incentive scheme \cite{TIS-104} securely rewards the intermediate nodes for relaying a message where a time order-preserving aggregated signature method is applied to detect the layer insertion attack. MuRIS \cite{IncentSharing-106} applies a rule to thwart the edge insertion attack in which the reward for relaying a message through an n-hop path must be equal or higher than the total rewards gained via an insertion attack.

\subsubsection{Miscellaneous Credit Approaches}
in addition to the game-theoretic and security-based mechanisms, some miscellaneous credit-based schemes have been proposed in the literature. Guan \textit{et al.} \cite{InterThreat-103} address the appearance of poverty nodes in DTNs in which SS nodes with strong social relations preferably choose each other as intermediate nodes to forward their messages, hence they gain more credit. Thus, it becomes difficult for nodes with fewer social relations to get sufficient rewards so that they can afford the cost of their forwarding messages. To deal with this problem, a taxation strategy is employed to fairly redistribute the credit among nodes and avoid the existence of poverty nodes. Mei and Stefa \cite{Give2Get-118} propose Give2Get Epidemic and Give2Get Delegation protocols based on a cryptographic proof-based technique to stimulate SS nodes to relay messages received from their non-social nodes. CAIS \cite{CAIS-109} aims to stimulate IS and SS nodes to participate in message relaying by designing different charging and rewarding strategies for IS and SS nodes. In particular, social and non-social credit are rewarded to a node when it relays a message received from a node in the same community or other communities, respectively. Furthermore, a data replication controlling mechanism is designed, based on which the number of messages a node can replicate is limited based on its social and non-social credit.

Seregina \textit{et al.} \cite{OnDesignIncent-110} design a reward-based incentive scheme for the Two-hop routing where the source of a message rewards only the first deliverer. Thus, intermediate nodes decide whether or not to relay a message based on the information provided by the source node. Specifically, three strategies are analyzed where the source can share information about the message in three settings: full (the number and ages of the message copies), partial (the number of the message copies), and no information. The experimental results reveal that the expected reward paid by a resource node is the same irrespective of the information provided to the relay node. Meanwhile, it is optimal for the source node to pretend that it is the first message replicator. DISCUSS \cite{DistCoop-114} is an incentive-based data forwarding protocol based on evolutionary theory in which encountered nodes share their message forwarding history with each other that can help them to choose the best routing strategy dynamically. In DISCUSS, three types of nodes are considered: cooperators that relay messages for others altruistically; exploiters that use the capability of other nodes in data forwarding but do not relay their messages; and isolators that neither help nor get help. Thus, a node in DISCUSS selects the cooperators as the next message carriers as well as motives the exploiters and isolators to reveal their routing strategies and cooperate in message relaying.

\begin{figure}[!t]
\centering
\includegraphics[width=3.4 in]{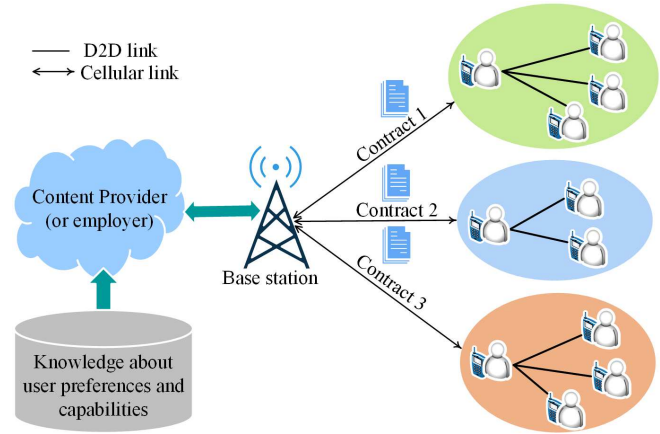}
\caption{A general illustration of contract-based incentive design.}
\label{ContractTheory}
\end{figure}

\begin{table*}[!t]
%% increase table row spacing, adjust to taste
%\renewcommand{\arraystretch}{1.3}
% if using array.sty, it might be a good idea to tweak the value of
% \extrarowheight as needed to properly center the text within the cells
\caption{Summary of the security-based and miscellaneous credit mechanisms}
\label{SecurityMiscellaneouscredits}
\centering
\linespread{1.1}\selectfont
%% Some packages, such as MDW tools, offer better commands for making tables
%% than the plain LaTeX2e tabular which is used here.
\begin{tabular}{|*{12}{c|}} % repeats {c|} X times
\hline

\multirow{3}{*}{\rotatebox[origin=c]{90}{\textbf{Model}}} & 	
%\multirow{3}{*}{\rotatebox[origin=c]{90}{\textbf{Ref.}}} &	
\multirow{3}{*}{\parbox[t]{13mm}{\textbf{Reference}}} &
\multirow{3}{*}{\parbox[t]{40mm}{\textbf{Principle of proposed solutions}}} &
\multirow{3}{*}{\parbox[t]{38mm}{\textbf{Incentive objective}}} &
\multirow{3}{*}{\parbox[t]{45mm}{\textbf{Specialties (+) and limitations (-)}}}  \\

\multicolumn{2}{|c|}{}&&& \\ \multicolumn{2}{|c|}{}&&& \\

\toprule
%-----------------------------------------

\hline
\parbox[t]{2mm}{\multirow{14}{0.2in}{\rotatebox[origin=c]{90}{Security-based credit approaches}}} &
\multirow{3}{0.7in}{Zhu \textit{et al.} \cite{SMART-100}}&	
\multirow{3}{1.7in}{A multi-layer credit scheme based on layered coin}&	
\multirow{3}{1.6in}{Dividing the total credit among the cooperative nodes based on a profit-sharing model}&	
\multirow{3}{2.1in}{+ Detects the edge insertion and removal attacks}  \\ &&&& \\ &&&& \\  \cline{2-5}
&
\multirow{3}{0.7in}{Lu \textit{et al.} \cite{Pi-102}}&
\multirow{3}{1.7in}{A hybrid (credit and reputation) incentive that provides fairness}&	
\multirow{3}{1.6in}{Maximizing the credit of nodes that deliver messages}&	
\multirow{3}{2.1in}{+ Thwarts edge insertion and removal attacks
\\- Considers the single-copy routing only}  \\ &&&& \\ &&&& \\  \cline{2-5}
&
\multirow{3}{0.7in}{Chen \textit{et al.} \cite{SecureCredit-107}}&
\multirow{3}{1.7in}{An incentive-compatible scheme for the nodes that have a finite budget}&	
\multirow{3}{1.6in}{Rewarding the nodes in the earliest delivery path based on the concept of contribution time}&	
\multirow{3}{2.1in}{+ Detects the edge insertion, removal, and manipulation attacks
\\- No evaluation of the communication cost}  \\ &&&& \\ &&&& \\ \cline{2-5}
&
\multirow{3}{0.7in}{Zhou and Cao \cite{TIS-104}}&
\multirow{3}{1.7in}{A threshold incentive mechanism based on a modified population dynamic model}&	
\multirow{3}{1.6in}{Rewarding the nodes for data relaying and security  considerations}&	
\multirow{3}{2.1in}{+ Considers fairness by providing equal relaying opportunities to each node
\\- No evaluation of the communication cost}  \\ &&&& \\ &&&& \\ \cline{2-5}
&
\multirow{3}{0.7in}{Wang \textit{et al.} \cite{IncentSharing-106}}&
\multirow{3}{1.7in}{A multi-receiver charging and rewarding scheme for data dissemination}&	
\multirow{3}{1.6in}{Replicating messages to nodes that have already delivered previous messages successfully}&	
\multirow{3}{2.1in}{+ Detects the edge insertion attacks
\\- Cannot detect colluding attacks}  \\ &&&& \\ &&&& \\  \cline{2-5}

%------------------------------------------------------------------
%----------------------------------------------------------------
\hline
\parbox[t]{2mm}{\multirow{35}{0.2in}{\rotatebox[origin=c]{90}{Miscellaneous credit approaches}}} &
\multirow{3}{0.7in}{Guan \textit{et al.} \cite{InterThreat-103}}&	
\multirow{3}{1.7in}{A taxation strategy to avoid the existence of poverty nodes caused by socially selfish behavior}&	
\multirow{3}{1.6in}{Provides credit for socially isolated nodes to afford buy the forwarding services of other nodes}&	
\multirow{3}{2.1in}{+ Introduces a new form of internal threats
\\- Lack of analytical evaluations}  \\ &&&& \\ &&&& \\  \cline{2-5}
&
\multirow{3}{0.7in}{Mei and Stefa \cite{Give2Get-118}}&
\multirow{3}{1.7in}{One of the first selfish-resilient social-aware data forwarding protocols}&	
\multirow{3}{1.6in}{Pushing messages far from a local community swiftly with a minimum number of replications}&	
\multirow{3}{2.1in}{+ The routing strategies are Nash equilibria
\\Cannot detect colluding selfish nodes}  \\ &&&& \\ &&&& \\ \cline{2-5}
&
\multirow{3}{0.7in}{Ning \textit{et al.} \cite{CAIS-109}}&
\multirow{3}{1.7in}{A community-based incentive scheme to stimulate both IS and SS nodes in data relaying}&	
\multirow{3}{1.6in}{Maximizing the individual and the social utility of nodes}&	
\multirow{3}{2.1in}{+ Applies different types of credit for nodes with different routing preferences
\\- Lack of analytical evaluations}  \\ &&&& \\ &&&& \\  \cline{2-5}
&
\multirow{4}{0.7in}{Seregina \textit{et al.} \cite{OnDesignIncent-110}}&
\multirow{4}{1.7in}{A credit scheme to promote the cooperation of nodes in Two-hop relaying}&	
\multirow{4}{1.6in}{Minimizing the amount of prices to be paid for delivering messages}&	
\multirow{4}{2.1in}{+ Every relay node is proposed a different reward based on its contact probability
\\- Rewarding is unfair because only the first deliverer is rewarded}  \\ &&&& \\ &&&& \\  &&&& \\ \cline{2-5}
&
\multirow{4}{0.7in}{Misra \textit{et al.} \cite{DistCoop-114}}&
\multirow{4}{1.7in}{A rewarding scheme in which nodes adapt their forwarding strategy based on message delivery information}&	
\multirow{4}{1.6in}{Maximizing the delivery probability of messages by motivating selfish nodes to cooperation}&	
\multirow{4}{2.1in}{+ Presenting both analytical and simulation-based experiments
\\- No consideration of the nodes' social preferences}  \\ &&&& \\ &&&& \\ &&&& \\ \cline{2-5}
&
\multirow{3}{0.7in}{Zhuo \textit{et al.} \cite{IncentOff-TMC}}&
\multirow{3}{1.7in}{An auction-based incentive mechanism that leverages nodes' delay tolerance for traffic offloading}&	\multirow{3}{1.6in}{Minimizing the incentive cost given an offloading target}&	
\multirow{3}{2.1in}{+ Considers the dynamic features of nodes' delay tolerance
\\- Nodes' social features are not considered
}  \\ &&&& \\ &&&& \\ \cline{2-5}
&
\multirow{3}{0.7in}{Li \textit{et al.} \cite{ContractOffload-TWC}}&
\multirow{3}{1.7in}{A contract-based incentive mechanism for data offloading with respect to nodes' satisfaction factors}&	
\multirow{3}{1.6in}{Maximizing the operator's profit for both continuous and discrete user-type models}&	
\multirow{3}{2.1in}{+ The operator can make decision based on nodes' statistical information
\\- Node mobility is not considered
}  \\ &&&& \\ &&&& \\ \cline{2-5}
&
\multirow{3}{0.7in}{Kouyoumdjieva and Karlsson \cite{EnergyOffload}}&
\multirow{3}{1.7in}{An adaptable and scalable mobile data offloading protocol under full and limited node cooperation}&	
\multirow{3}{1.6in}{Maximizing the network throughput while saving the nodes' energy}&	
\multirow{3}{2.1in}{+ exploits the energy consumption of nodes that participate in the offloading process
\\- Does not conduct analytical experiments
}  \\ &&&& \\ &&&& \\ \cline{2-5}
&
\multirow{3}{0.7in}{Zhang \textit{et al.} \cite{ContractIncentive2015}}&
\multirow{3}{1.7in}{A contract-based mechanism to overcome the information asymmetry problem in D2D content sharing}&	
\multirow{3}{1.6in}{Optimizing the network capacity while guaranteeing the network QoS requirements}&	
\multirow{3}{2.1in}{+ A flexible rewarding method based on the nodes' preferences
}  \\ &&&& \\ &&&& \\ \cline{2-5}
&
\multirow{4}{0.7in}{Chen \textit{et al.} \cite{ContractIncentive2017}}&
\multirow{4}{1.7in}{A general framework for designing optimal contracts between the operator and D2D nodes}&	
\multirow{4}{1.6in}{Maximize the profit of service provider and nodes according to their valuations}&	
\multirow{4}{2.1in}{+ The operator does not require gathering information from nodes frequently
\\- Less communication and computational costs}  \\ &&&& \\ &&&& \\  &&&& \\ \cline{2-5}
&
\multirow{3}{0.7in}{Zhao \textit{et al.} \cite{FusingSocial-115}}&
\multirow{3}{1.7in}{A social and contact-based incentive scheme for community-based D2D data sharing}&	
\multirow{3}{1.6in}{Maximizing the utility of nodes and their social friends with respect to their restricted resources}&	
\multirow{3}{2.1in}{+ Selfish nodes are stimulated to truthfully report their data forwarding preferences
}  \\ &&&& \\ &&&& \\ \cline{2-5}
&
\multirow{3}{0.7in}{Pan \textit{et al.} \cite{PreferWill2017}}&
\multirow{3}{1.7in}{A social-based incentive scheme for community-based D2D data offloading}&	
\multirow{3}{1.6in}{Maximizing the data offloading gain considering the nodes' content and social preferences}&	
\multirow{3}{2.1in}{+ Complimenting simulations with analytical results
}  \\ &&&& \\ &&&& \\ \cline{2-5}
&

\multirow{4}{0.7in}{Wu \textit{et al.} \cite{SocialRate2017}}&
\multirow{4}{1.7in}{A social-aware rate-based D2D data sharing scheme, which is modeled as a maximum weighted mixed matching problem}&	\multirow{4}{1.6in}{Maximizing the individual utility of nodes}&	
\multirow{4}{2.1in}{+ Considers a novel multi-hop D2D communication paradigm
\\- Resource representations and scheduling techniques are not considered}  \\ &&&& \\ &&&& \\ &&&& \\

\hline
%\multicolumn{12}{c}{(``$\surd$'' if the protocol satisfies the property, ``$\times$'' if not)}
\end{tabular}
\end{table*}

Some recent incentive-based mobile data offloading mechanisms aim at encouraging mobile nodes to relay a portion of the cellular traffic through DTNs and Wi-Fi hotspots \cite{MobOffliadingSur}. Zhou \textit{et al.} \cite{IncentOff-TMC} propose a reverse auction-based incentive approach, namely Win-Coupon, in which nodes with high delay tolerance and large offloading potentials have the highest priority to offload the cellular traffic. In Win-Coupon, auction-winning users receive data with delay and earn a coupon, whereas other nodes download data from the cellular network directly. In particular, a semi-Markov model is designed to predict the nodes' delay tolerance potentials based on their mobility patterns. Similarly, Li \textit{et al.} \cite{ContractOffload-TWC} employ contract theory to model delayed data offloading between an operator and mobile nodes in which each mobile node chooses a proper contract based on its preferences. The main objective is to maximize the operator's profit while guaranteeing the feasibility of the nodes. The authors in \cite{EnergyOffload} propose an energy-aware mobile data offloading algorithm, which combines duty cycling and selfishness energy saving mechanisms to promote the cooperation of mobile nodes. The experiments reveal that the proposed scheme achieves up to 85\% energy savings while losing about 1\% in system throughput when nodes fully cooperate in data distribution. In addition, it shows that the proposed scheme is robust against non-cooperative nodes even when 50\% of the nodes do not follow the underlying data offloading protocol.

%Similarly, Li \textit{et al.} \cite{YongLiOffload-TMC} model mobile data offloading as a submodular function maximization problem and propose three solutions

A number of miscellaneous incentive approaches have been proposed in D2D communication. The authors in \cite{Contract3}\cite{ContractIncentive2015}\cite{ContractIncentive2017} introduce the application of contract theory to model the interaction between content provider(s) and nodes where the main objective is to maximize the utility of the operator provided that the expected utility of nodes is also satisfied when signing the contract. In comparison to other incentive approaches (\textit{e.g.}, auction games), contract-based methods can reduce nodes' computational and communication cost because the operator does not need to collect the nodes' feedback after each auction announcement. Instead, the operator provides different contracts and their corresponding rewards for nodes with different features, and the nodes can select a more beneficial contract with maximum benefits (Fig. \ref{ContractTheory}).

Some incentive schemes for D2D communications group nodes into communities based on their social relationships or contact history and explore their incentives for inter and intra-group cooperations. Zhao \textit{et al.} \cite{FusingSocial-115} propose a three-phase approach for data dissemination in which nodes are grouped into communities based on their betweenness centrality. Next, seed nodes in each community are identified according to their closeness centrality. Finally, they disseminate messages received from the BS to their socially-connected nodes in their community where the nodes in each community have an incentive to mutually benefit from exchanging messages with each other in a multi-hop D2D communication mode.
Similarly, Pan \textit{et al.} \cite{PreferWill2017} propose a content pushing mechanism in which nodes are grouped based on their content preferences where a node replicates contents for inter-group and intra-group nodes with different probabilities. The experiments demonstrate that the offloading performance heavily relies on the cooperation level of nodes. Wu \textit{et al.} \cite{SocialRate2017} propose a joint social-aware and link quality-based content sharing mode selection protocol in the presence of cooperative and SS nodes. It is assumed that there exist three communication models: BS-to-D2D, D2D, and multi-hop D2D. Thus, the content sharing mode selection problem is modeled as a maximum mixed matching problem.

\textbf{Summary:} Table \ref{SecurityMiscellaneouscredits} summarizes the main characteristics of the security-based and miscellaneous credit mechanisms. It can be seen that the majority of the security-based rewarding schemes use the layered coin technique to protect granting rewards to cooperative nodes and protect the rewarding system against malicious attacks. In addition, almost all the security-based credit schemes focus on DTNs, while the security of credit distribution in D2D communications is not studied in the existing works. Furthermore, the miscellaneous credit mechanisms employ concepts, such as taxation, contract theory, and social community to design their incentive mechanisms.

\noindent
\section{Open Discussion and Future Directions}
\label{FutureWork}
In previous sections, we have reviewed the state-of-the-art of data routing and dissemination services and protocols in the non-cooperative WRNs and highlighted their specialties and limitations. In light of the works focusing on various aspects of the non-cooperative WRNs, there are still several open problems and challenges, which are left without proper answers. In this section, we discuss possible future research directions that can bring new visions into the horizon of WRNs.

\noindent
\subsection{Realistic Human Altruism and Selfishness Models}
So far, we introduced different types of human selfish behaviors and actions in WRNs (\textit{e.g.}, \cite{Selfishalturisim-14,EmpiricalAlturistic-13}). Although IS and SS nodes have been introduced as general selfishness models, several other important factors (such as available resources, content knowledge, and spatiotemporal information) should be further explored to realistically model the selfish behavior of mobile nodes in WRNs. For example, it is challenging how the selfish behavior of mobile nodes evolves in different situations and locations based on their social and contextual properties. In addition, it is non-trivial to explore how the selfish behavior of nodes changes when different levels of battery or power resources remain in their devices (or when their devices are charging). Modeling human selfish behaviors in D2D communication with respect to its unique characteristics is another challenging issue that received less attention from the research community. For example, it is not explored how much selfish D2D nodes have the freedom to limit sharing their spectrum resources with other nodes. Moreover, how their social tie information and relationships affect their cooperation levels in content sharing and distribution.

\noindent
\subsection{Impact Analysis of Human Non-cooperative Behaviors on Data Forwarding and Content Sharing}
Although the impact of mobile nodes' non-cooperative behaviors on the performance of data delivery protocols in DTNs has been studied from different perspectives (see Section \ref{Impact}), several avenues for further research are still open. The existing analytical models have generally explored the effects of nodes' selfish behavior on only the data delivery delay and transmission cost metrics (see Table \ref{tableImpact}). One future trend is extending the existing analytical frameworks to a generic model (\textit{e.g.}, a multi-dimensional CTMC model) to analyze the performance depredation of other system parameters (such as the data delivery ratio and energy) and compare their tradeoffs. In addition, exploring the impact of nodes' sophisticated selfish behavior on the overall performance of data delivery raises new research problems. For example, it is non-trivial to explore how nodes' social ties, physical locations, or contextual information affect their cooperation level and the performance of data delivery protocols under different settings (\textit{e.g.}, when the network traffic varies from medium to high).

The impact of mobile nodes' selfish behavior on the overall D2D network performance is another interesting research challenge that received less attention by the research community. Although a limited number of simulation-based experiments (\textit{e.g.}, \cite{SocialD2DOffload}\cite{SelfishD2DYongLi}) have studied human selfish behaviors in D2D communications underlying cellular networks, there is no analytical approach to explore the effects' of node selfishness on network throughput accurately. For example, a CTMC model can be designed to model data dissemination in community-based D2D communications and analyze how the network performance metrics are degraded in the presence of D2D selfish mobile nodes. Additionally, specific communication protocols and policies (\textit{e.g.}, opportunistic scheduling algorithms) should be developed to determine human cooperation models and control the system parameters against changes made by D2D selfish mobile nodes.

\noindent
\subsection{Robust Mechanisms to Detect Non-cooperative Nodes}
Although different mechanisms are proposed to detect selfish and malicious mobile nodes in WRNs (see Section \ref{Detection}), they might be ineffective and inefficient in case the number of malicious nodes is high or sophisticated denial-of-service attacks are launched by them. The main reason is that mobile nodes often do not have up-to-date information about the network's global state (\textit{i.e.}, the contact and social graphs), especially in highly dynamic WRNs. One promising solution to effectively detect non-cooperative nodes is establishing trust relationships among nodes based on their social similarities or analyzing their data forwarding behaviors based on their social preferences (\textit{e.g.}, see \cite{SoWatch}). This idea sounds very useful because the social features of nodes are relatively stable over time. Another possible solution is developing a learning system based on nodes' contact history or social relations to discover the patterns of common selfishness and attack models. In such a system, mobile nodes can upload their contact and social properties to a server and the server runs complicated operations to learn the nodes' behavior and find their selfishness and attack patterns. Detecting colluding attackers in WRNs is another challenging problem, which is addressed by a few numbers of recent studies (\textit{e.g.}, in \cite{ColluBlakGrey-34}). While the majority of the existing detection methods investigate nodes' contact history to discover inconsistent or manipulated records, exploring the contact graph (instead of contact history) can help detect colluding attackers swiftly and accurately.

Establishing secure and reliable data sharing and dissemination protocols in D2D communications by selecting honest and trustworthy intermediate nodes and isolating selfish and malicious nodes is a greatly challenging problem. For example, it is non-trivial to explore how to detect D2D selfish nodes in heterogeneous and large-scale networks when they use unlicensed bands to share their messages. One promising solution could be to design distributed and decentralized security and trustworthy mechanisms in which novel technologies (such as blockchain) are employed to store and exchange nodes' security information and control their cooperation and trustworthiness. Another exciting research direction is to detect malicious nodes and their attack models in D2D communications, which is not explored in the literature.

\noindent
\subsection{Effective and Fair Incentive Mechanisms}

Different incentive mechanisms have been proposed to stimulate the cooperation of selfish nodes in WRNs (see Section \ref{Incentive}). Overall, it can be seen that effectiveness and fairness are two important factors that should be considered in designing an incentive mechanism. In other words, an incentive scheme should not only appropriately encourage selfish nodes to help relay messages on behalf of other nodes but also reward the nodes according to their cooperation level fairly and protect the rewarding system against malicious attacks and unfair manipulations. One major challenge in providing effective incentives is to devise various forms of incentives (\textit{e.g.}, monetary, social relevance, or non-monetary) to stimulate the cooperation of nodes with different selfish behaviors and preferences. For example, empirical experiments in \cite{EmpiricalAlturistic-13} reveal that minor credit (\textit{e.g.}, one dollar) can change the altruistic behavior of mobile users with limited device resources significantly. In addition, taking into account the properties of contents (\textit{e.g.}, their size) and the actual capabilities of nodes for data distribution (\textit{e.g.}, the energy level of their devices) can help design effective incentive mechanisms.

Another challenging future research direction is developing effective incentive strategies in D2D-enabled heterogeneous networks that can ultimately raise cooperation among wireless D2D nodes, spectrum owners, and service providers. Due to the bandwidth limitations of the backhaul network and base stations in the heterogeneous networks, encouraging resource-limited D2D devices to cache contents for others and share their spectrum resources with them is extremely challenging, especially for data-intensive applications with massive users. To effectively stimulate D2D devices to cooperate with the other network entities in data delivery, different criteria, such as resource availability and user interests in content should be considered. For example, different cost and rewarding mechanisms can be considered for users with different preferences. One promising solution is applying cooperative and non-cooperative game-theoretic approaches to analyze multi-stage interactions between the base station and D2D devices with heterogeneous resources, reveal their true preferences, and maximize their utilities.

Developing secure and privacy-preserving incentive mechanisms in the presence of malicious and cheating nodes are other important research challenges that need further explorations. For instance, how to design secure incentive mechanisms that enforce the required fairness in assigning rewards to cooperative nodes is still an open problem. Besides, it remains an important issue how to stimulate mobile nodes to cooperate in data delivery, consume their computational, and bandwidth resources while preserving their privacy.

\section{Conclusion}
\label{Conclusion}
In this paper, we presented an in-depth literature review of recent studies on human-centric communications in non-cooperative WRNs. Specifically, we introduced different selfish behavior and malicious attacks that can be launched by misbehaving nodes in cooperative data delivery. Meanwhile, we studied the impacts of nodes' different non-cooperative actions on the performance of data delivery protocols. In addition, we discussed distributed detect and defense mechanisms that attempt to identify selfish and malicious nodes in WRNs. Furthermore, we explored a large number of incentive mechanisms and discussed their major characteristics. Finally, we discussed several open issues and future research directions. Since efficient and secure communications are simultaneously becoming ever-important in next-generation wireless networks, we hope that this survey will be useful for the network protocol and mobile application developers and encourage them to design appealing data delivery mechanisms.

\section*{Acknowledgments}
This work is partially supported by the National Natural Science Foundation of China (61572106 and 61502075). The authors are grateful to the anonymous reviewers for their constructive comments and suggestions to improve the quality of the article.

\bibliographystyle{IEEEtran}
\bibliography{Survey}

\end{document}